\documentclass[journal=jctcce,manuscript=article]{achemso}
\usepackage[utf8]{inputenc}

\setkeys{acs}{usetitle = true}
\setkeys{acs}{maxauthors=10,etalmode=truncate}
\usepackage{xspace, amsmath, multirow, setspace, graphicx, subfigure}
\usepackage{bm, dcolumn, epsfig, color, xr, booktabs, longtable, lscape, titlecaps}
\Addlcwords{the of into to in for a an with on and at from by}
\mciteErrorOnUnknownfalse
\usepackage{cleveref} 
\usepackage{amsmath}

\title[]{Local Hydration Control and Functional Implications Through
  S-Nitrosylation of Proteins: Kirsten rat sarcoma virus (KRAS) and
  Hemoglobin (Hb)}

\author{Haydar Taylan Turan$^{1}$ and Markus Meuwly$^{1}$}
\affiliation[misc]{ $^{1}$Department of Chemistry, University of
  Basel, Klingelbergstrasse 80, Basel, Switzerland}
\email{m.meuwly@unibas.ch}


\begin{document}
\doublespace

\maketitle
\thispagestyle{empty}

\begin{abstract}
S-nitrosylation, the covalent addition of NO to the thiol side chain
of cysteine, is an important post-transitional modification (PTM) that
can affect the function of proteins. As such, PTMs extend and
diversify protein functions and thus characterizing consequences of
PTM at a molecular level is of great interest. Although PTMs can be
detected through various direct/indirect methods, they lack the
capabilities to investigate the modifications at the molecular
level. In the present work local and global structural dynamics, their
correlation, the hydration structure, and the infrared spectroscopy
for WT and S-nitrosylated Kirsten rat sarcoma virus (KRAS) and
Hemoglobin (Hb) are characterized from molecular dynamics
simulations. It is found that for KRAS attaching NO to Cys118
rigidifies the protein in the Switch-I region which has functional
implications, whereas for Hb nitrosylation at Cys93 at the $\beta_1$
chain increases the flexibility of secondary structural motives for Hb
in its T$_{0}$ and R$_{4}$ conformational substates. Solvent water
access decreased by 40\% after nitrosylation in KRAS, similar to Hb
for which, however, local hydration of the R$_4$NO state is yet lower
than for T$_0$NO. Finally, S-nitrosylation leads to detectable peaks
for NO stretch, however, congested IR region will make experimental
detection of these bands difficult
\end{abstract}

\today

\maketitle

\section{Introduction}
Post-translational modifications (PTM) comprise covalent
1modifications of proteins that occur at one or several
sites\cite{walsh2006posttranslational} which can lead to structural,
local hydration changes and can alter the physio-chemical properties
of the proteins and binding partners.\cite{scott2009cell} They can
impact protein function both allosterically and
orthosterically.\cite{nussinov2012allosteric} PTMs can be divided into
two main groups. The first group classifies the covalent additions of
a - usually - electrophilic fragment of a cosubstrate to the side
chains of the amino acid\cite{walsh2005protein} whereas the second
group classifies the cleavage of peptide backbones by autocatalytic
cleavage or proteases.\cite{walsh2012three} Nitrosylation,
phosphorylation, oxidation, alkylation, alkylation, and glycosylation
are among the most common types of covalent side-chain
modifications.\cite{walsh2005protein} Backbone PTMs range from
methylation of amide nitrogen\cite{van2017autocatalytic} to
significant alterations to the backbone.\cite{burkhart2017ycao}\\

\noindent
A wide range of experimental techniques has been proposed and applied
to detect the PTMs. These methods can be divided into two groups:
indirection detection usually breaks the covalent bond between
modification and protein and then captures the signals of the PTMs
adduct part whereas direct techniques address the adducts
directly. Indirect methods such as
biotin-switch\cite{jaffrey2001protein}, Enzyme-Linked immunosorbent
Assay (ELISA)\cite{voller1978enzyme} are robust, but they usually
require multi-step sequential manipulation which may hamper the
selectivity and reproducibility.\cite{alcock2018chemical} Furthermore,
they usually only provide whether or not PTM has occurred at a
particular site but do not allow for further in-depth physicochemical
or structural characterization of the consequences of PTM. Direct
methods such as erasable single-molecule blotting (eSiMBlot) or
organophosphine probes are subject to similar selectivity problems as
indirect methods as well.\cite{alcock2018chemical} Further,
spectroscopic techniques can be utilized for the detection of PTM. The
mass spectroscopy is one of the widely used techniques for PTM
detection.\cite{doll2015mass}. Nuclear magnetic resonance (NMR)
spectroscopy is used to detect PTM and characterize their effects on
protein conformation.\cite{kumar2019characterizing}. Finally, infrared
(IR) spectroscopy, is one of the most direct methods as it reports
directly on the bond themselves.\\

\noindent
The human genome is estimated to comprise $\sim$20.000 to 25.000 genes
whereas the human proteome is estimated at over 1 million
proteins.\cite{mclaughlin2016and}. The diversity of proteome is
provided by two major mechanisms. The first one is mRNA splicing at
the transcriptional level, and the second one are PTMs. Hence, PTMs
help to extend and diversify proteins function, and in turn the
diversity of organism, further than gene transcripts
dictate.\cite{wang2014protein}. So, it is important to characterize
the effects of PTMs at an atomistic level to have a better
understanding about the alteration in protein function induced by
modifications ranging from regulation of cell
survival\cite{case2011mechanical}, protein aggregation via covalent
cross-linking\cite{davies2016protein} to protein
unfolding.\cite{hagai2010ubiquitin} \\
   
\noindent
Kirsten rat sarcoma virus (KRAS) is one of the three members of the
rat sarcome virus (RAS) oncogene family along with the HRAS and NRAS
proteins that plays a role in human cancer.\cite{simanshu2017ras} Over
20\% of human cancers contain mutated RAS genes which makes them the
most frequent oncogenic drivers\cite{colicelli2004human}, while KRAS
is accounted for 85\% of RAS mutations.\cite{reck2021targeting} The
pancreatic (88\%), colorectal (50\%) and lung cancers (35\%) are types
of human cancer with the highest rate of KRAS
mutations.\cite{prior2020frequency} KRAS protein switches between
guanosine diphosphate (GDP)-bound inactive state and guanosine
triphosphate (GTP)-bound active state. The transformation from a
stable GDP-bound state to the active GTP-bound state is stimulated by
guanine nucleotide exchange factor (GEF) proteins. The transformation
back to the inactive GDP-bound state is mediated by GTPase-activating
proteins (GAP). The switch between active and inactive states is
highly regulated and responsive to multiple signal inputs due to the
fact that the switch is controlled by GEFs and
GAPs.\cite{simanshu2017ras}.  \\

\noindent
Hemoglobin (Hb) is an important and extensively studied protein due to
its role in transporting oxygen from the lungs to the tissues. Deoxy
(T$_0$) and oxy (R$_4$) structures are two of the most important
structural states of the Hb.\cite{baldwin1979haemoglobin} The
stability of deoxy and oxy states is subject to the number of ligands
bound to the heme-iron. The deoxy state is stable when no ligand is
bound to the ferrous iron atom of heme whereas the oxy state is stable
when four ligands are bound. Beside oxygen several other ligands such
as carbon monoxide (CO), carbon dioxide (CO$_2$) and nitric oxide (NO)
can be bonded to the iron as well. Hb can bind carbon monoxide to the
iron and form carboxyhemoglobin which inhibits the oxygen binding due
to the occupation of the binding site. Further, binding affinity of
hemoglobin is $\sim 250$ times higher than its affinity for the
oxygen. \\
  
\noindent
Protein hydration and protein-water interactions are crucial for chain
folding, structure, and conformational stability of
proteins.\cite{denisov1996protein} Further, it is known that
interactions between proteins and individual water molecules can
mediate protein functions such as recognition, binding, or
catalysis.\cite{halle2004protein} Although water molecules can also
occupy internal cavities and deep clefts, the majority of protein
hydration studies focuses on the water interactions with the external
surface of the protein.\cite{edsall1983water, olson2007,
  chen2008hydration, schiro2019role }. The water molecule layer which
in intimate interaction with a protein surface is called protein
hydration shell.\cite{moron2021protein}. Local hydration is important
in that it can inhibit or promote a PTM by altering solvent and
cofactor accessibility around the surrounding region of the
modification site. The local hydration of the protein also can be
modified by PTM. S-nitrosylation is known to alter the water structure
around the cysteine residue.\cite{turan2021spectroscopy} Oxidation of
Met to the sulfoxide MetSO renders the side chain of Met both, polar
and hydrophilic.\cite{hardin2009coupling} The phosphorylated
kinase-induced domain (pKID) showed more hydrophobic interactions
after including the modification which promotes the formation of the
special hydrophobic residue cluster which includes residues Leu128,
Tyr134, Ile137, Leu138, and Leu141.\cite{liu2020phosphorylation}
Moeller et al. discussed the phosphorylation-dependent regulation of
AQP2 water permeability. Also, they state ``phosphorylation can change
hydrophobic regions of a protein into polar (negatively charged) or
hydrophilic''.\cite{moeller2011regulation} Although the importance of
local hydration has been pointed out in the literature, no direct
proof of local hydration modification has been reported
experimentally.\\

\noindent
The present work is structured as follows; First, atomic simulations
and methods used in the paper are described. This is followed by the
discussion of hydration structure, and local hydrophobicity around the
modification site. Then, structural effects and dynamical coupling
induced by S-nitrosylation are discussed. Finally, the IR spectra of
WT and modified proteins are presented, and conclusions are drawn.

\section{Computational Methods}
\subsection{Molecular Dynamics}
All molecular Dynamics (MD) simulations were performed using the
CHARMM\cite{charmm.prog} software with the
CHARMM36\cite{huang2013charmm36} force field. The equations of motion
were propagated with a leapfrog integrator\cite{hairer2003geometric},
using a time step of $\Delta t = 1$ fs and all bonds involving
hydrogen atoms were constrained using
SHAKE.\cite{ryckaert1977numerical} Non-bonded interactions were
treated with a switch function\cite{steinbach1994new} between 12 and
16 \AA\/ and electrostatic interactions were computed with the
particle mesh Ewald method.\cite{darden1993particle}\\

\noindent
For the simulations involving wild-type and S-nitrosylated KRAS,
twelve different simulations were set up: wild-type KRAS (PDB:
4OBE)\cite{hunter2014situ} at 50 and 300 K, wild-type KRAS without GDP
at 50 and 300 K, cis- and trans-S-nitrosylated KRASNO at 50 and 300 K,
and cis- and trans-S-nitrosylated KRASNO without GDP at 50 and 300
K. Two chains with identical sequence were available in the crystal
structure. Chain A was selected for the simulation setup. The
cis-KRASNO and trans-KRASNO set up starting from the same initial
structure except for the dihedral angle $\phi$(C$_{\beta}$SNO)
(0$^{\circ}$ for cis- and 180$^{\circ}$ for trans-KRASNO). \\

\noindent
For the simulations involving wild-type and S-nitrosylated Hb twenty
different simulations were set up: wild-type T$_0$ Hb (PDB: 2DN2) at
50K and 300 K, wild-type R$_4$ Hb (PDB: 2DN3) at 50K and 300K, cis-
and trans-S-nitrosylated T$_0$ Hb (T$_0$NO) at 50 and 300 K with PC
and MTP, cis- and trans-S-nitrosylated R$_4$ Hb (R$_4$NO) at 50 and
300 K with PC and MTP. cis- and trans- conformers are set up starting
from the same initial structure except for the dihedral angle
$\phi$(C$_{\beta}$SNO) (0$^{\circ}$ for cis- and 180$^{\circ}$ for
trans-KRASNO).\\

\noindent
KRAS and Hb were solvated in a $66 \times 66 \times 66$ \AA\/$^3$ and
$90 \times 90 \times 90$ \AA\/$^3$ cubic boxes of
TIP3P\cite{jorgensen1983comparison} water, respectively.  The protein
was weakly constrained to the middle of the simulation box, minimized,
heated to the desired temperature, and equilibrated for 500 ps in the
$NVT$ ensemble. Production runs of 10 ns were then performed in the
$NpT$ ensemble using the leapfrog Verlet
integrator\cite{verlet1967computer} ($\Delta t$ = 1 fs) and a Hoover
barostat\cite{hoover1985canonical} with a collision rate of 5
ps$^{-1}$. The force field employed for the -SNO moiety was
parameterized as described in the previous
study.\cite{turan2021spectroscopy} \\

\subsection{Infrared Spectroscopy}
The molecular dipole moment ($\mu$) of the protein was calculated from
the MD trajectories and the partial charges. Subsequently, the Fourier
transform of the dipole moment autocorrelation function computed to
obtain the infrared spectrum. The autocorrelation function
\begin{equation}
\label{correlation}
C(t) = \langle\overrightarrow{\mu}(0) \cdot
\overrightarrow{\mu}(t) \rangle
\end{equation}
was accumulated over $2^{16}$ time origins to cover 1/3 to 1/2 of the
trajectory. From this, the absorption spectrum is determined according
to
\begin{equation}
\label{ir}
A(\omega) = \omega (1 - e^{-h \omega/(k_{\rm B} T)}) \int C(t)
e^{-i \omega t} dt
\end{equation}
where $T$ is the temperature in Kelvin, $k_{\rm B}$ is the Boltzmann
constant, and the integral is determined using a fast Fourier
transform (FFT). IR spectra of WT KRAS, cis-KRASNO, and trans-KRASNO,
WT Hb, cis-T$_0$NO, trans-T$_0$NO, cis-R$_4$NO and trans-R$_4$NO have
been generated for blocks of 100 ps simulation by correlating over
$2^{16}$ time origins. A total of 100 spectra were generated for each
system (total simulation time of 10 ns) and averaged.\\

\noindent
In addition to IR spectra, the power spectrum of the NO bond was
calculated from the FFT of the bond length time-series autocorrelation
function to provide assignments of the vibrational spectra and allows
to detect couplings between modes.\cite{lammers2007investigating}
These power spectra were not averaged and correlated over $2^{16}$
time origins for the entire simulation time of 10 ns.\\

\subsection{Dynamical Cross-Correlation Maps}
The dynamical cross-correlation maps (DCCM) and difference dynamical
cross-correlation maps \cite{ichiye1991collective,
  arnold1997molecular} ($\Delta$DCCM) were calculated to
quantitatively characterize the effects of S-nitrosylation on the
protein dynamics using the Bio3D package.\cite{grant2006bio3d} Dynamic
cross-correlation maps matrices and coefficients
\begin{equation}
    \label{dccm_func}
    C_{ij} = \langle\Delta r_{i} \cdot \Delta r_{j}\rangle /  (\langle\Delta r_{i}^{2} \langle\Delta r_{j}^{2}\rangle)^{1/2}
\end{equation} 
were determined from the position of C$_{\alpha}$ in amino acids $i$
and $j$ with positions $r_{i}$ and $r_{j}$. $\Delta r_{i}$ and $\Delta
r_{j}$ determine the displacement of the $i$th C$_{\alpha}$ from its
average position throughout the trajectory. One should note that DCCM
characterizes the correlated ($C_{ij} > 0$) and anti-correlated
($C_{ij}$ $<$ 0) motions in a protein whereas $\Delta$DCCM reports on
the distinct differences between unmodified and modified protein.

\section{Results}
\begin{figure}[H]
  \centering \includegraphics[width=0.45\linewidth]{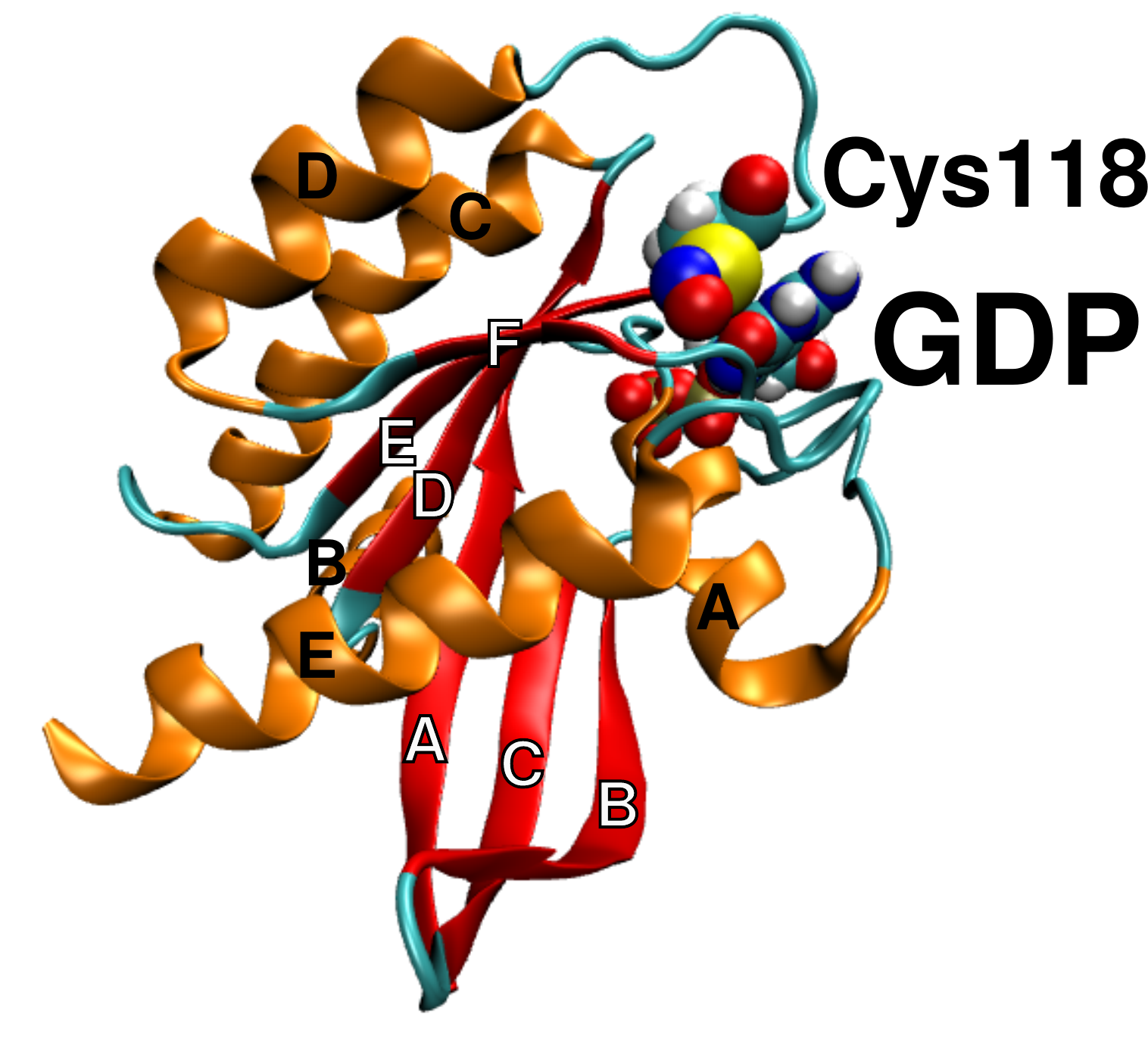}
  \includegraphics[width=0.48\linewidth]{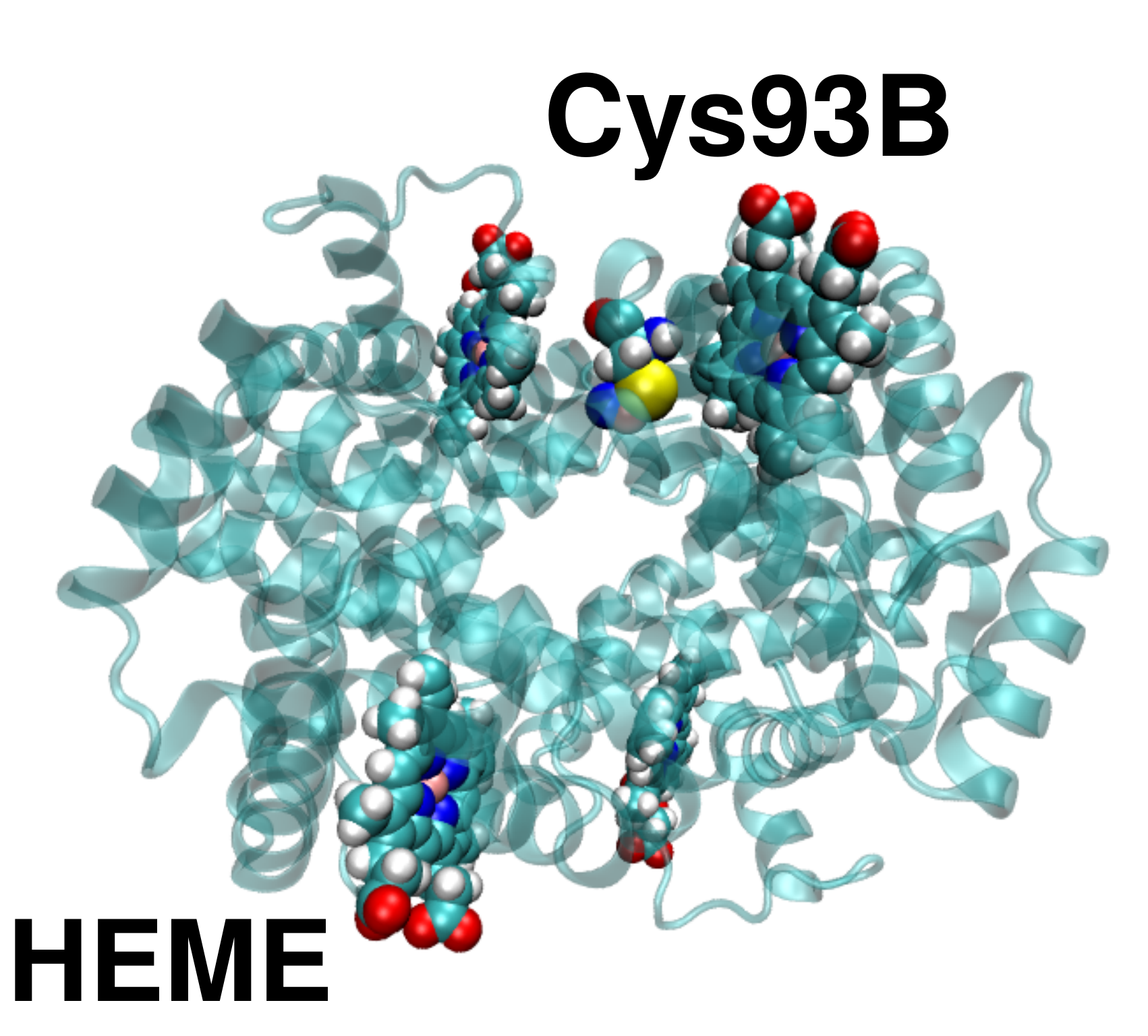}
\caption{Left: Structure of trans-KRASNO with GDP (PDB: 4OBE). The
  labeled $\alpha$-helices are shown in orange, $\beta$-sheets are in
  red and loops are in cyan. Right: Structure of cis-T$_0$NO (PDB:
  2DN2). The S-Nitrosylated Cys118 and Cys93$\beta$ residues, GDP and HEME
  are represented by CPK.}
\label{fig1}
\end{figure}

The present work reports on the solvation, structural, dynamical and
spectroscopic implications of S-nitrosylation at Cys118 for KRAS and
at Cys93$\beta$ for Hb, see Figure~\ref{fig1}.

\subsection{Local and Global Structural Changes}
\begin{figure}[H]
  \centering
    \includegraphics[width=0.75\linewidth]{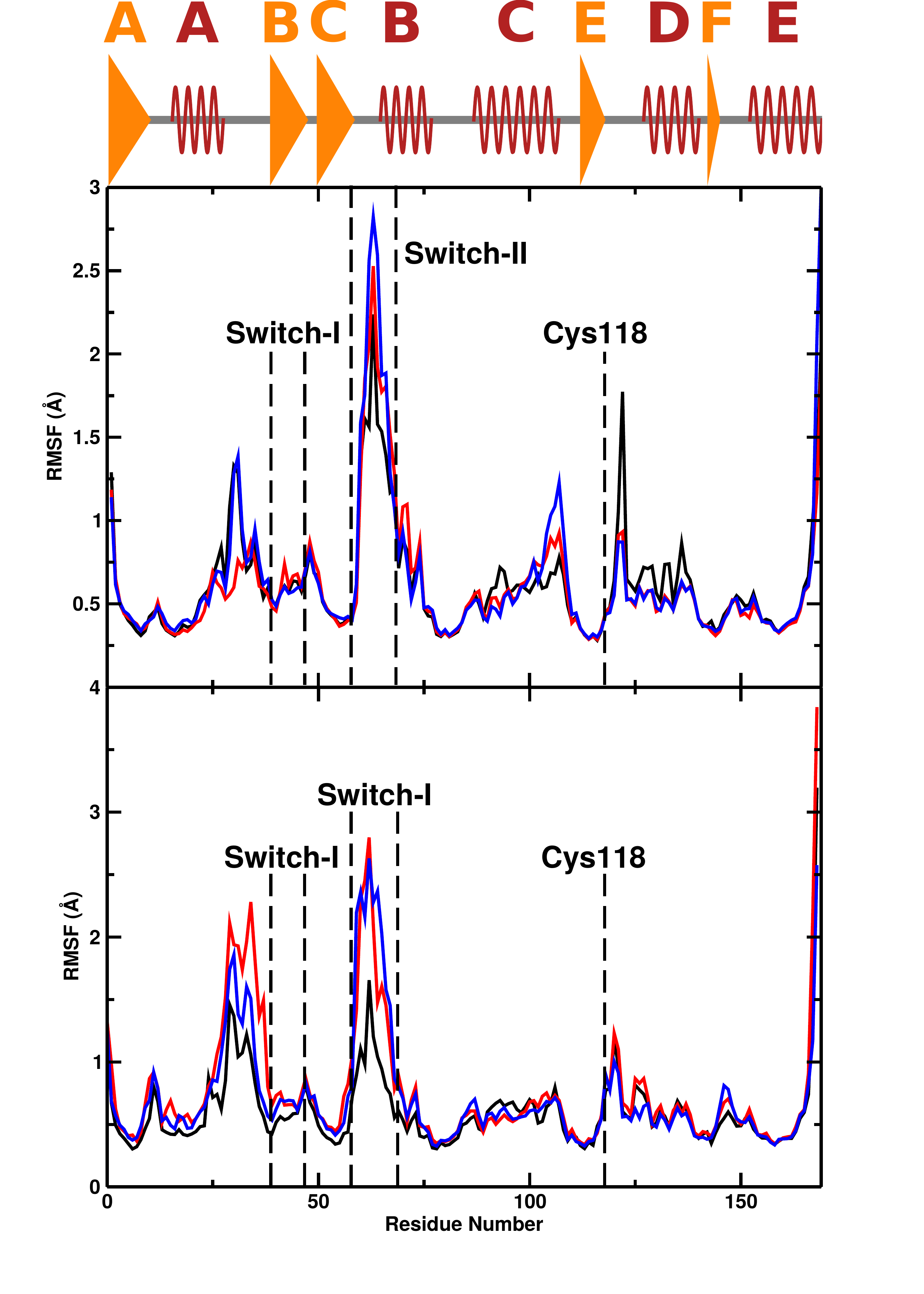}
    \caption{Root Mean Square Fluctuation of each residue at 300 K
      from simulations with (top) and without GDP (bottom) using the
      PC model for WT (black), cis-KRASNO (red), trans-KRASNO
      (blue). Orange triangles indicate the position of $\beta$-sheets
      and red helices indicate the position of $\alpha$-helices.}
\label{fig2}
\end{figure}

{\it KRAS:} The root-mean-squared fluctuations of the C$_{\alpha}$
atoms of every residue at 300 K from 10 ns simulations with and
without GDP are reported in Figure~\ref{fig2} (top and bottom) for WT
(black), cis-KRASNO (red), and trans-KRASNO (blue). For simulations
with GDP, attaching NO decreased the flexibility of residues Phe28 to
Asp33 (Loop B) for cis-KRASNO compared to WT and trans-KRASNO. The
importance of this region is arising from the fact the residues are on
the Switch-1 region (Phe28 to Asp38). Switch regions act as a binding
interface for effector proteins and RAS
regulators.\cite{pantsar2020current} Therefore, rigidification of
these regions potentially affect the binding characteristics of KRAS
protein. Moreover, the second switch region (Switch-II, Tyr58 to
Tyr64) is also affected by the modification. The modification
increased the flexibility of residues Tyr58 to Arg68 in both cis- and
trans-KRASNO with respect to WT. The RMSF values increased up to 2.8
\AA\/ and 2.5 \AA\/ for trans- and cis-KRASNO, respectively, from 1.6
\AA\/ for WT.  Increased flexibility was also observed for the
residues Val103 to Asp108 which are resides on the end of Helix C and
the beginning of Loop D. Interestingly, the flexibility of Ser122 was
significantly higher in WT compared to nitrosylated variants.\\

\noindent
For the simulations without GDP, NO attachment induced increased
flexibility of both Switch regions. The RMSF values increased for
residues His27 to Ile45, the residues reside on Switch-1 (Loop B) and
$\beta$-sheet B. However, unlike simulations with GDP, both cis- and
trans-KRASNO had increased flexibility in this region compared to
WT. The RMSF values increased up to 2.3 and 1.9 \AA\/ for cis- and
trans-KRASNO, respectively. Further, the flexibility increased also
observed in the Switch-II region from residues Tyr58 to Asp69. The
RMSF values increased to 2.8 and 2.6 \AA\/ for cis- and trans-KRASNO,
and comparable to those from simulations with GDP in the active
site. S-nitrosylation had a marginal effect on the
flexibility/rigidification of the rest of the residues. Also, the
flexibility of loop E (Cys118 to Asp126) where Cys118 resides was less
affected from the NO attachment compared to Loop B and Loop C
regions.\\

\noindent
The calculated RMSF values at 300 K superimposed onto the
experimentally measured C$_{\alpha}$ B-factors and results are shown
in Figure~\ref{sifig1}. For direct comparison, the B-Factor and RMSF
values are scaled and normalized. The computed RMSF values for WT
agree qualitative with the experimental B-factors except for Switch-II
and Helix B region (Tyr58 to Gly77).  The agreement between experiment
and simulation remains consistent until His27 which is the end residue
of Helix A. Similar to WT simulations, an agreement was observed
between the experiment and simulations for cis- and trans-KRASNO
except for Switch-II and Helix B region. The agreement again remains
consistent until the end of helix A. Additionally, agreement from the
beginning of Sheet E to the C-terminus of the protein is found for the
nitrosylated variants. The lack of agreement for the Switch-II and
Helix B region could be due to the fact that two chains are present in
the crystal structure compared to using a single chain in the
simulations. The differences between experiments and simulations also
may arise from effects such as crystal packing or lattice disorder
which would not be present in simulations.\cite{sun2019utility}\\

\begin{table}[H]
\centering
\caption{The average C$_{\alpha}$ RMSD (in \AA\/) for Sheet D, Sheet
  E, Sheet F, Helix C, Helix D, Helix E, Loop B, Loop E and the entire
  protein for WT, cis-KRASNO and trans-KRASNO with respect to the WT
  X-Ray structure for the last nanosecond of a 10 ns free dynamics
  simulation at 300 K with PC.}
\begin{tabular}{l | c  | c  | c}

\hline
 &  WT & cis & trans \\
\hline
Sheet D   &0.47  &0.57  &0.56 \\
Sheet E   &0.36  &0.60  &0.43 \\
Sheet F   &0.46  &0.66  &0.40 \\
Helix C   &0.99  &1.15  &1.07 \\
Helix D   &0.77  &0.95  &0.70 \\
Helix E   &0.91  &1.45  &1.87 \\
Loop B    &0.89  &4.42  &1.17 \\
Loop E    &1.53  &0.85  &0.72 \\
Protein   &1.28  &1.75  &1.53 \\
\hline
\label{tab1}
\end{tabular}
\end{table}

\noindent
The last nanosecond of the 10 ns production run was analyzed to
characterize the structural changes at 300 K. Even though the
structural changes induced by the modification were evident for the
entire protein structure to a certain degree, eight regions revealed
eminent changes. These are: sheet D [Ser65,Glu76], sheet E (Val112 to
Lys117), sheet F (Ile142 to Tyr144), helix C (Tyr87 to Ser106), helix
D (Tyr127 to Ile139), helix E (Val152 to Lys169), loop B (Ser39 to
Asp47) and Loop E (Cys118 to Asp126). The average C$_{\alpha}$ RMSD
(in \AA\/) for these 8 regions and the entire protein for WT,
cis-KRASNO and trans-KRASNO with respect to the WT X-Ray structure for
the last nanosecond of a 10 ns free dynamics simulation at 300 K with
PC are summarized in Table~\ref{tab1} for simulations
with GDP.  \\

\noindent
The effect of the modification to the global structure of the protein
was more evident at 300 K, as expected. The RMSD increased by 0.47 and
0.25 \AA\/ for cis- and trans-KRASNO with respect to WT,
respectively. However, the most dramatic increase in C$_{\alpha}$ RMSD
was observed for loop B in cis-KRASNO, see Section
\ref{chap:dccm}. The displacement increased to 4.42 \AA\/ from 0.89
\AA\/ in WT. Since, loop B is also the Switch-I part of the protein,
this increase in displacement after S-nitrosylation can have a
significant effect on the binding interface characteristics of KRAS to
effector proteins and RAS regulators. Also, loop E, where Cys118
resides, is rigidified by the modification the RMSD values are
decreased in half for cis- and trans-KRASNO compared to WT.\\

\begin{table}[h]
\centering
\caption{The average C$_{\alpha}$ RMSD (in \AA\/) for helix F, helix
  G, helix H, loop B, loop E and chain $\beta_1$ for cis-T$_0$NO,
  trans-T$_0$NO, T$_0$, cis-R$_4$NO, trans-R$_4$NO, R$_4$ with respect
  to the T$_0$ and R$_4$ X-Ray structures for the last nanosecond of a
  10 ns free dynamics simulation at 300 K with PC and MTP models.}
\begin{tabular}{l | c c  | c c | c | c c | c c | c}
\hline
 &  \multicolumn{2}{c}{cis-T$_0$NO}& \multicolumn{2}{c}{trans-T$_0$NO}& T$_0$ & \multicolumn{2}{c}{cis-R$_4$NO}& \multicolumn{2}{c}{trans-R$_4$NO} & R$_4$\\
\hline
 & PC & MTP & PC  & MTP & PC & PC & MTP & PC & MTP & PC \\
helix F  &1.19  &1.23 &1.23  &1.26 &1.08 &1.48  &1.49  &{\bf 1.71}  &{\bf 1.71} &{\bf 1.80}   \\
helix G  &1.30  &1.28 &1.47  &1.46 &0.98 &1.66 &1.68  &0.98  &0.99 &1.35  \\
helix H  &1.17  &1.16 &1.34  &1.26 &1.26 &1.15 &1.14  &1.24 &1.32 &1.42  \\
loop B   &{\bf 2.24}  &{\bf 2.18} &{\bf 1.97} &{\bf 2.00} &1.24 &0.96 &1.01  &0.96 &0.99 &0.88  \\
loop E  & 1.66 & 1.60 & 1.61 & 1.66 & 0.77 & 1.14 & 1.28 & 0.91 & 0.95 & 0.74 \\
Chain $\beta_1$  &0.98  &0.98 & 1.19 &1.21 & 0.93&0.95  &0.95  &1.04 &1.05 &0.80  \\
\hline
\label{tab2}
\end{tabular}
\end{table}

\noindent
{\it Hemoglobin}: The RMSD for the T$_{0}$ and R$_{4}$ structural
substates of hemoglobin with S-nitrosylation at Cys93$\beta$, located
on helix F, for both, the cis-, and trans-conformers was considered
next. The C$_{\alpha}$ atoms for cis-[T$_{0}$,R$_{4}$]NO (orange), and
trans-[T$_{0}$,R$_{4}$]NO (indigo) were superimposed on the
[T$_{0}$,R$_{4}$] (cyan) X-Ray structure at 300 K, see
Figure~\ref{sifig2}. The average C$_{\alpha}$ RMSD for helix F, helix
G, helix H, loop B, loop E and chain $\beta_1$ at 300 K are shown in
Table~\ref{tab2} for the T$_0$ and R$_4$ states. For T$_0$ the
C$_{\alpha}$ RMSD of Chain $\beta_1$ for cis- and trans-T$_0$NO with respect
to T$_0$ X-Ray structure are 0.98, 1.19 \AA\/ with PC and 0.98, 1.21
\AA\/ with MTP, respectively. For the secondary structural motifs, the
modification increased the C$_{\alpha}$ RMSD with respect to T$_0$
state for helix F, helix G, loop B and loop E. In general the RMSD
increases more for trans-T$_0$NO than for cis-T$_0$NO. The increase in
RMSD was marginal for helix G whereas significant increased have been
observed for helix G, loop B, loop E. The RMSD increased by 1 \AA\/ in
cis- conformer with respect to T$_0$ with PC. The RMSD of loop B was
2.24 \AA\/ in cis- conformer and 1.25 \AA\/ in T$_0$ with PC whereas
it was 2.18 \AA\/ for cis- with MTP. Also, significant increased have
been observed for the trans-T$_0$NO as well. The RMSD increased to
1.97 \AA\/ with PC and 2.00 \AA\/ with MTP. For loop E, RMSD also
increased $\sim 0.9$ \AA\/ for S-Nitrosylated variants with respect to
T$_0$ for both PC, and MTP. \\

\noindent
For the R$_4$ state, the C$_{\alpha}$ RMSD of Chain $\beta_1$ for cis- and
trans-R$_4$NO with respect to R$_4$ X-Ray structure are 0.98, 1.48
\AA\/ with PC and 1.48, 1.54 \AA\/ with MTP, respectively. The RMSD of
helix F and helix H decreased after the modification for both
conformers. For helix G, there was a significant increase in
cis-R$_4$NO with respect to R$_4$ with the values of 1.66 and 1.36
\AA\/, respectively, whereas trans-R$_4$NO rigidified, and RMSD
decreased to 0.982 with PC values. The loop B showed slightly more
displacement in both nitrosylated variants compared to R$_4$ state. No
significant deviation between PC and MTP C$_{\alpha}$ RMSD values have
been observed, the highest deviation was 0.08 \AA\/ increase for helix
H in trans-R$_4$NO.\\

\subsection{Dynamical Cross Correlation Maps}
\label{chap:dccm}
\begin{figure}[H]
  \centering
  \includegraphics[width=0.49\linewidth]{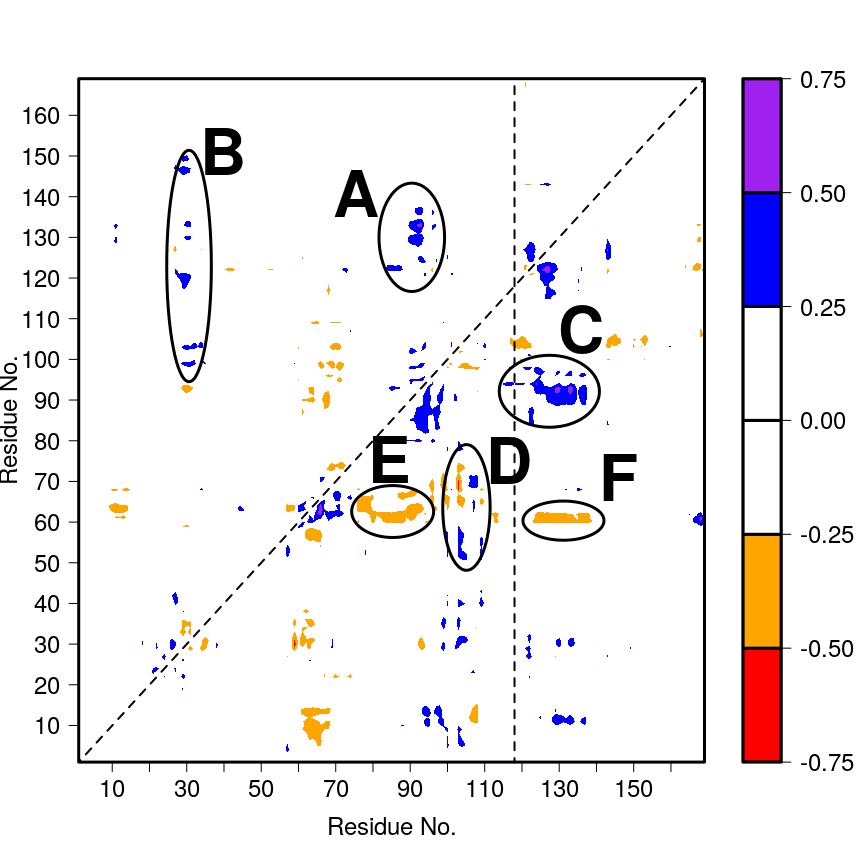}
  \includegraphics[width=0.49\linewidth]{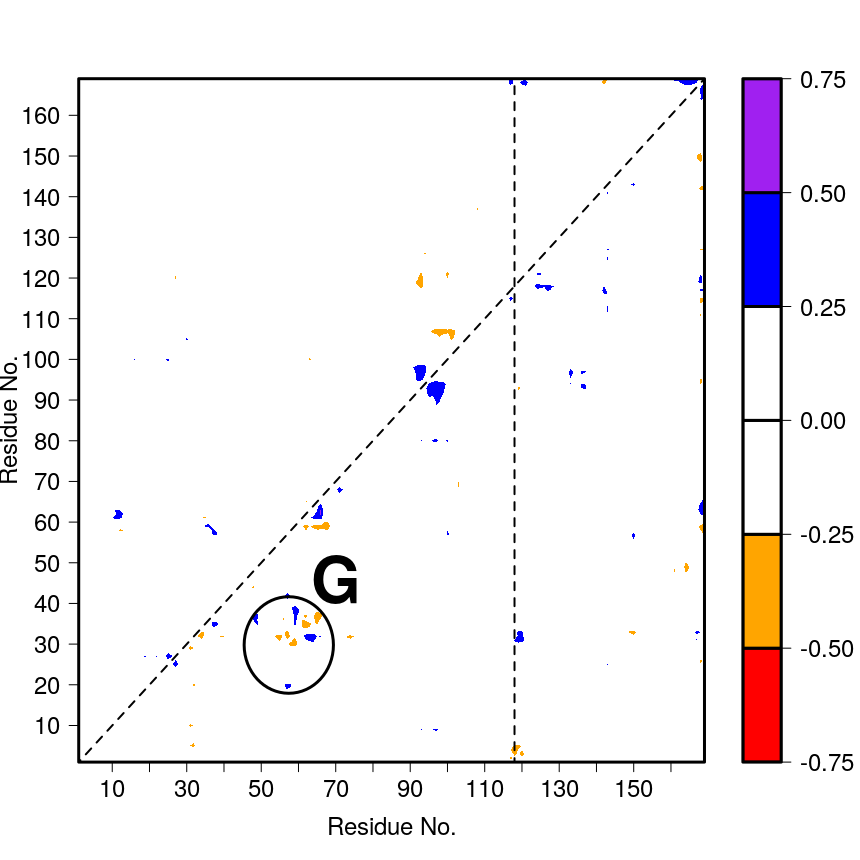}
  \includegraphics[width=0.49\linewidth]{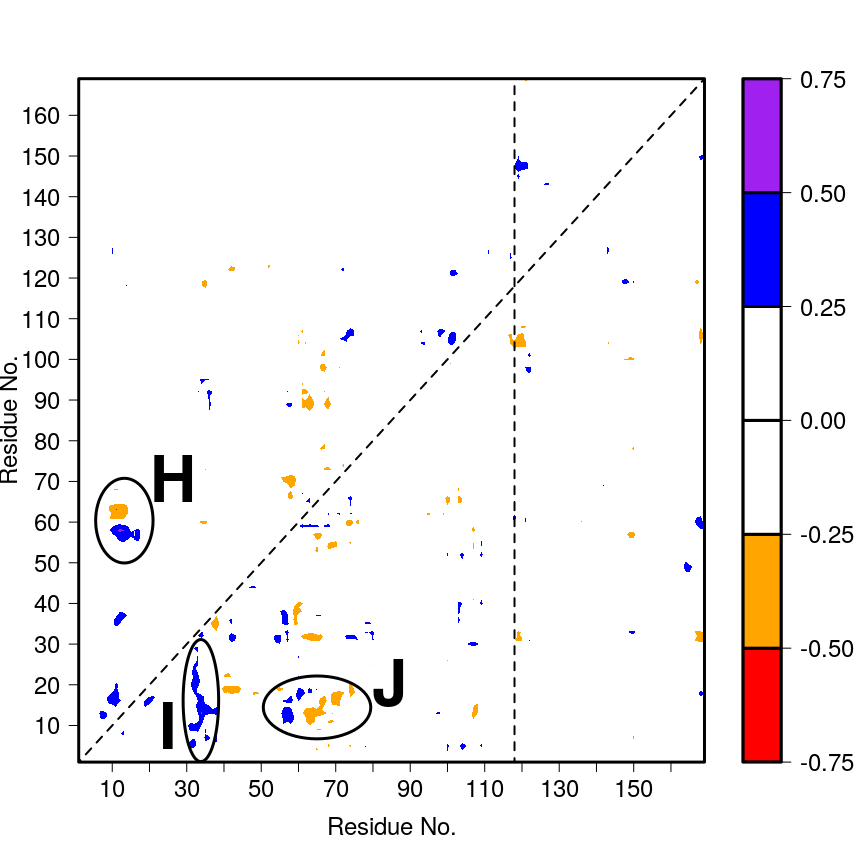}  \includegraphics[width=0.45\linewidth]{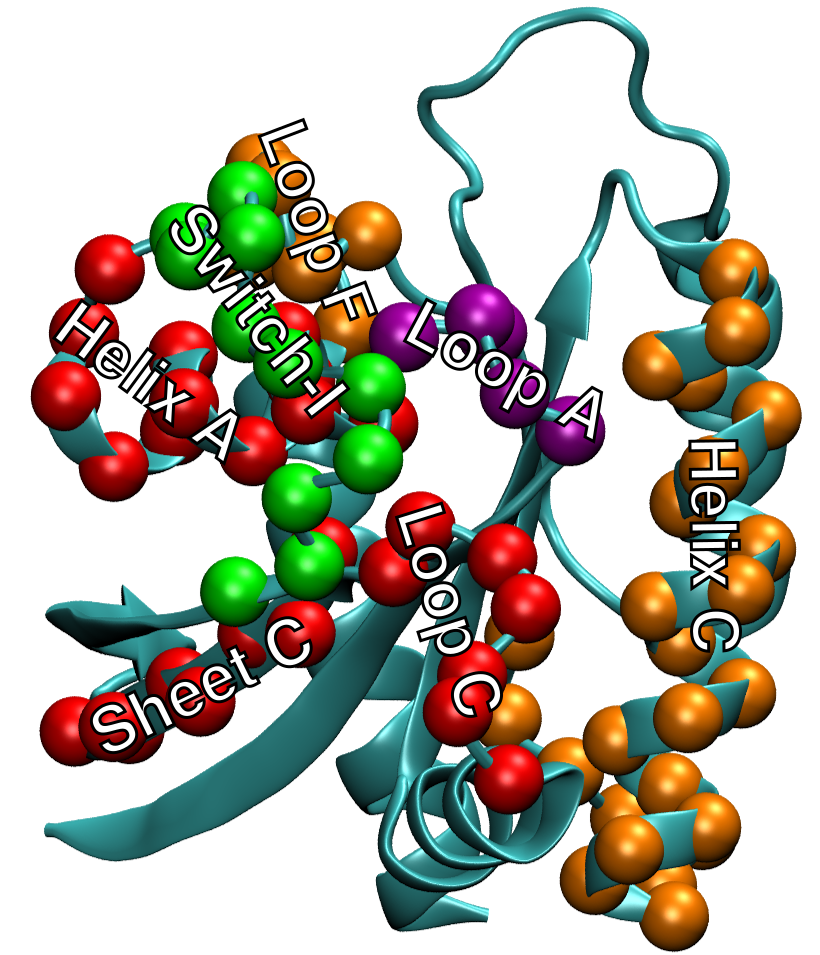}
  \caption{Top left panel: $\Delta$DCCM for GDP-bound KRAS from
    simulations at 300 K. Left upper triangle reports difference
    between cis-KRASNO and WT; right lower triangle reports difference
    between trans-KRASNO and WT. Top right panel:$\Delta$DCCM for
    ligand-free KRAS from simulations at 300 K. Left upper triangle
    reports difference between cis-KRASNO and WT; right lower triangle
    reports difference between trans-KRASNO and WT. The vertical and
    horizontal lines indicate the position of Cys118. Particular
    features in the maps are labeled from A to G with circles around
    them. Bottom left panel:$\Delta$DCCM for cis-KRASNO with and
    without GDP (left upper triangle) and trans-KRASNO with and
    without GDP (right lower triangle) at 300 K. The vertical and
    horizontal lines indicate the position of Cys118. Bottom right
    panel: Features B (Switch-I in green, and helix C loop D, loop F
    in orange) and H (Loop A in purple, and helix A, sheet C, loop C
    in red) visualized on KRAS.}
\label{fig3}
\end{figure}

\begin{figure}[H] \centering
    \includegraphics[width=0.49\linewidth]{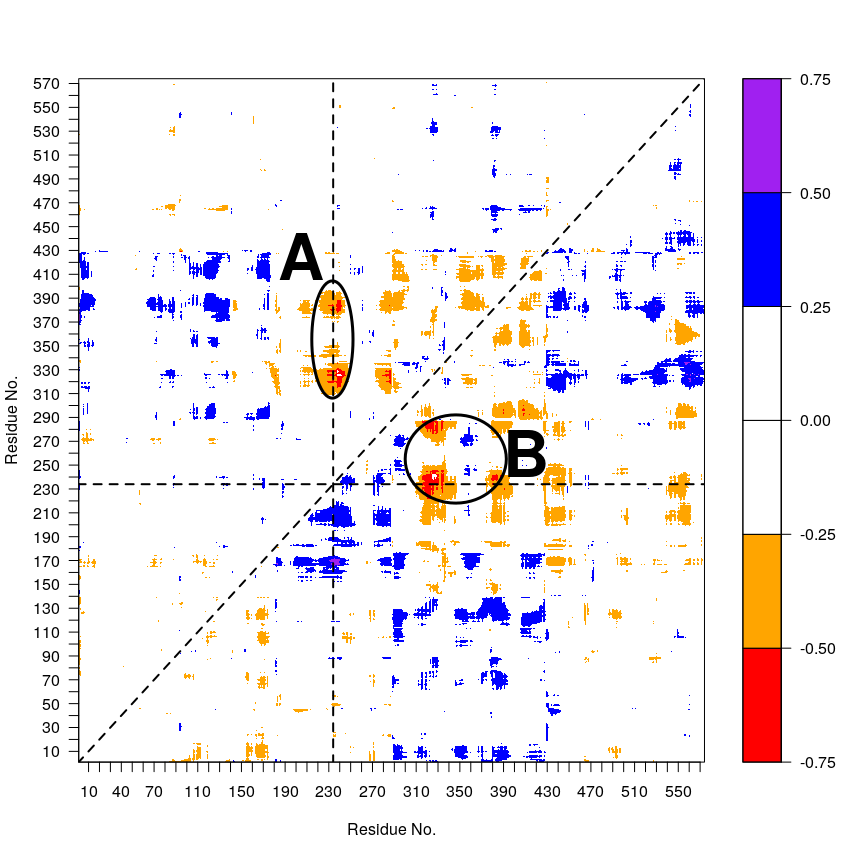}
    \includegraphics[width=0.49\linewidth]{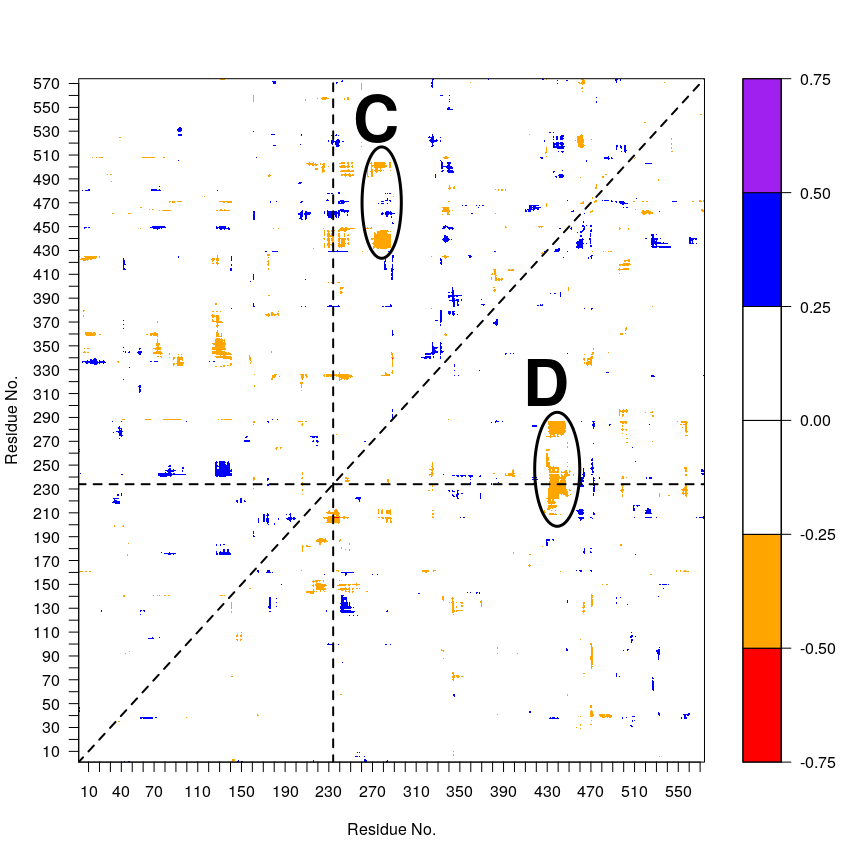}
    \caption{Left panel: $\Delta$DCCM for cis-T$_0$NO and T$_0$ (left
      upper triangle) and trans-T$_0$NO and T$_0$ (right lower
      triangle) at 300 K. Right panel: $\Delta$DCCM for cis-R$_4$NO
      and R$_4$ (left upper triangle) and trans-R$_4$NO and R$_4$
      (right lower triangle) at 300 K. Particular features in the maps
      are labeled from A to D with circles around them.}
\label{fig4}
\end{figure}

To further analyze the effects of -SNO modification, dynamical cross
correlation maps (DCCM) and their differences ($\Delta$DCCM) were
determined. Difference maps were calculated for cis-KRASNO
($\Delta$DCCM$_{\rm cis}$) and trans-KRASNO ($\Delta$DCCM$_{\rm
  trans}$) with WT as the reference and are shown in
Figure~\ref{fig3}. The WT DCCM was selected as a reference for
$\Delta$DCCM to report on pronounced differences between modified and
unmodified proteins and are presented in Figure~\ref{fig3}. For Hb,
$\Delta$DCCM for [cis, trans]-T$_0$NO with T$_0$ as the reference, and
similarly $\Delta$DCCM for [cis, trans]-R$_4$NO with R$_4$ as the
reference are presented in Figure~\ref{fig4}. \\

\noindent
{\it KRAS:} For the simulations without GDP, both $\Delta$DCCM$_{\rm
  cis}$ and $\Delta$DCCM$_{\rm trans}$ had limited number of
correlated motions between the different local parts of the protein
compared to the simulations with GDP, see Figure~\ref{fig3}. The
residues mainly had correlated movements within their local part with
adjacent residues, except for feature G which labels the correlated
motions between sheet B and helix B. \\

\noindent
For simulations with GDP, $\Delta$DCCM$_{\rm cis}$, feature A
corresponds to correlated movements between helix C and helix D with
0.25 $\leq$ $C_{ij}$ $\leq$ 0.75. Also, the correlated movement of the
Switch-I region with multiple regions such as helix C, loop D, and
loop F was observed in feature B. The results show that nitrosylation
not only increased the C$_\alpha$ RMSD of Switch-I but also its
displacement started to correlate with other local parts of the
protein. Especially, the correlated movement between the Switch-I
region and helix C shows that residues both can be correlated locally
or through space. For $\Delta$DCCM$_{\rm trans}$, correlated motion
between helix C and helix D was observed with higher intensity in
feature C. Helix C [Tyr87,Ser106] has correlated motions with loop F
[Glu107,Met111] (feature D) whereas Helix C has anti-correlated
motions with residues from sheet A [Met1,Gly10] to sheet D
[Phe78,Ile84]. Also, features E and F show that loop C is involved in
anti-correlated motions with residues [Gln70,Phe90] and
[Leu120,Pro140], respectively. \\

\noindent
To quantify the effects of the removal of GDP, $\Delta$DCCM are
calculated between cis-/trans-KRASNO with and without GDP, and the
results are shown in Figure~\ref{fig3}. For the cis-isomer, feature H
shows correlated motions between loop A with helix A and sheet C
whereas there are anti-correlated motions with loop C. Since the
diphosphate part of GDP resides near loop A and helix A, the motion of
these regions is important for the local dynamics of the
protein. Similar features but with larger correlation coefficients
were also observed for trans-KRASNO. Feature I shows prominent
correlated motion between sheet B and residues from Lys5 to Asp30
(sheet A, loop A, and helix A) whereas sheet C had correlated motions
with loop A and helix A and loop C showed anti-correlated motions with
these regions, see Feature J.\\

\noindent
Overall, the DCCM maps show that S-nitrosylation not only has a
significant effect on individual displacements of local structures but
also influence local correlated motions. Several correlated motion
features were observed in all systems, but they extend over up to 40
residues and the magnitude of the correlation increased after the
modification, especially for trans-KRASNO. Lastly, the removal of GDP
decreased the correlated motions significantly which emphasizes the
dynamical effects induced by ligand binding to the KRAS protein.\\

\noindent
{\it Hemoglobin:} $\Delta$DCCMs are calculated for cis-T$_0$NO and
T$_0$, trans-T$_0$NO and T$_0$, cis-R$_4$NO and R$_4$, trans-R$_4$NO
and R$_4$, and presented in Figure~\ref{fig4}. The structural motions
were significantly more correlated for T$_0$NO compared to
R$_4$NO. The anti-correlated motions which have $C_{ij} \leq -0.75$
were observed for both cis- and trans-T$_0$NO. Two particular features
are labeled in T$_0$, see left panel of Figure~\ref{fig4}. Feature A
shows the anti-correlated motions between [Leu87B,Leu115B] (including
the modified Cys93$\beta$ residue), and [Ala28C,Tyr42C] in
cis-T$_0$NO. Feature B shows strong anti-correlated motions between
[Ala28C,Leu48C] and [Asp79B,Arg104B] in trans-T$_0$NO. For R$_4$
proteins motions were less correlated and have values between -0.50
$\leq$ $C_{ij}$ $\leq$ 0.50. In feature C, there were anti-correlated
motions between [Thr4D,Glu22D] and [Ala70B,Leu114B] in
cis-R$_4$NO. Similarly, feature D showed anti-correlated motions
between [Ala129B,His146B] and [Leu3D,Val20D] in
trans-R$_4$NO. Overall, the correlated motions were mostly between
chain $\beta_1$ (including Cys93NO) and chain C. The results emphasize the
effect of the modification on the correlated motions of Hb protein.\\

\subsection{Hydration Structure Around the Modification Site}
Next, local hydration around the S-nitrosylated Cys were analyzed by
means of radial distribution function, and their corresponding
coordination numbers for both KRAS, and Hb. Given the evident role
that water can play for the protein function (see Introduction), such
a change in hydration may also be functionally relevant for a PTM such
as S-nitrosylation.\\

\begin{figure}[H]
  \centering
    \includegraphics[width=0.49\linewidth]{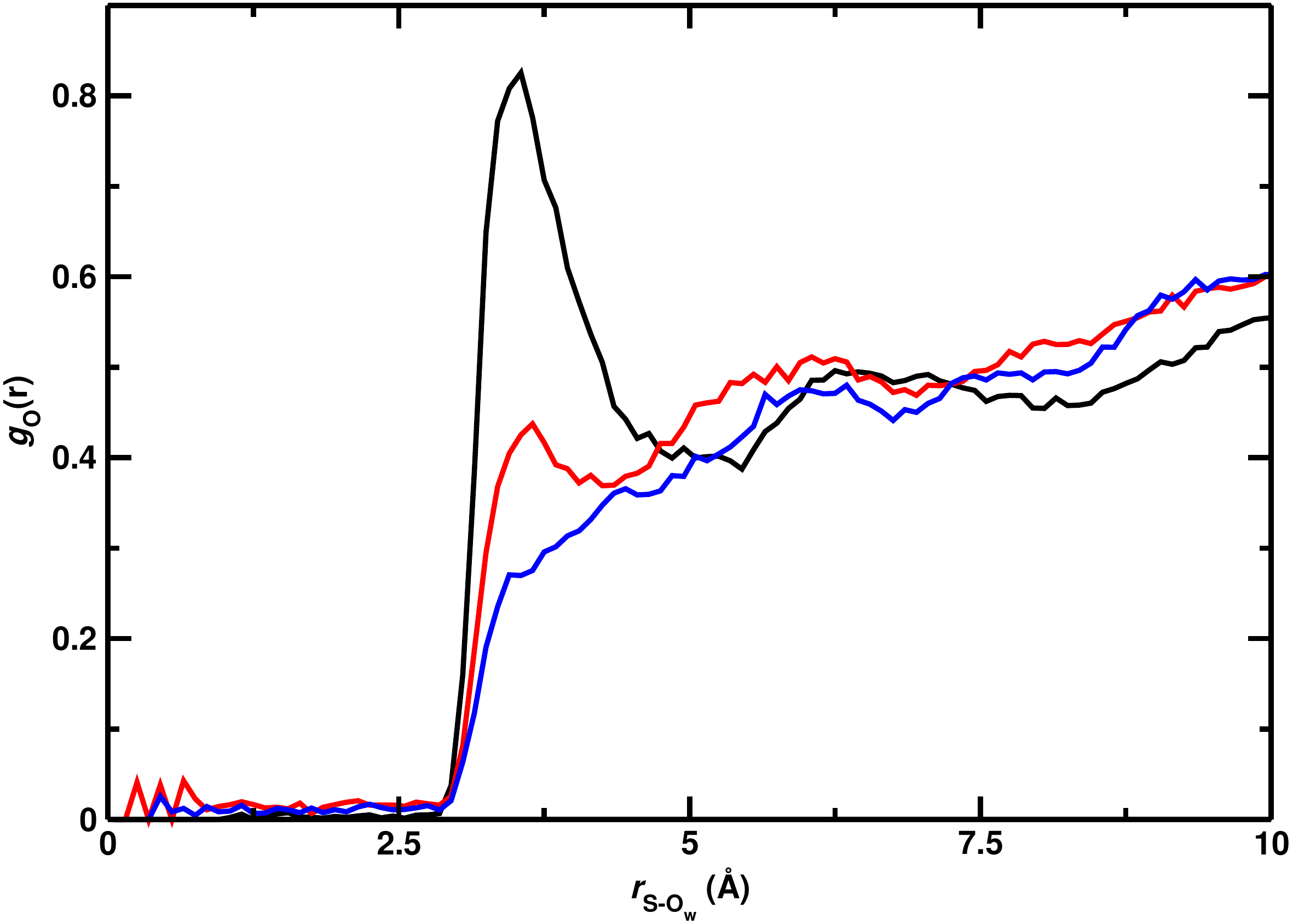}
    \includegraphics[width=0.49\linewidth]{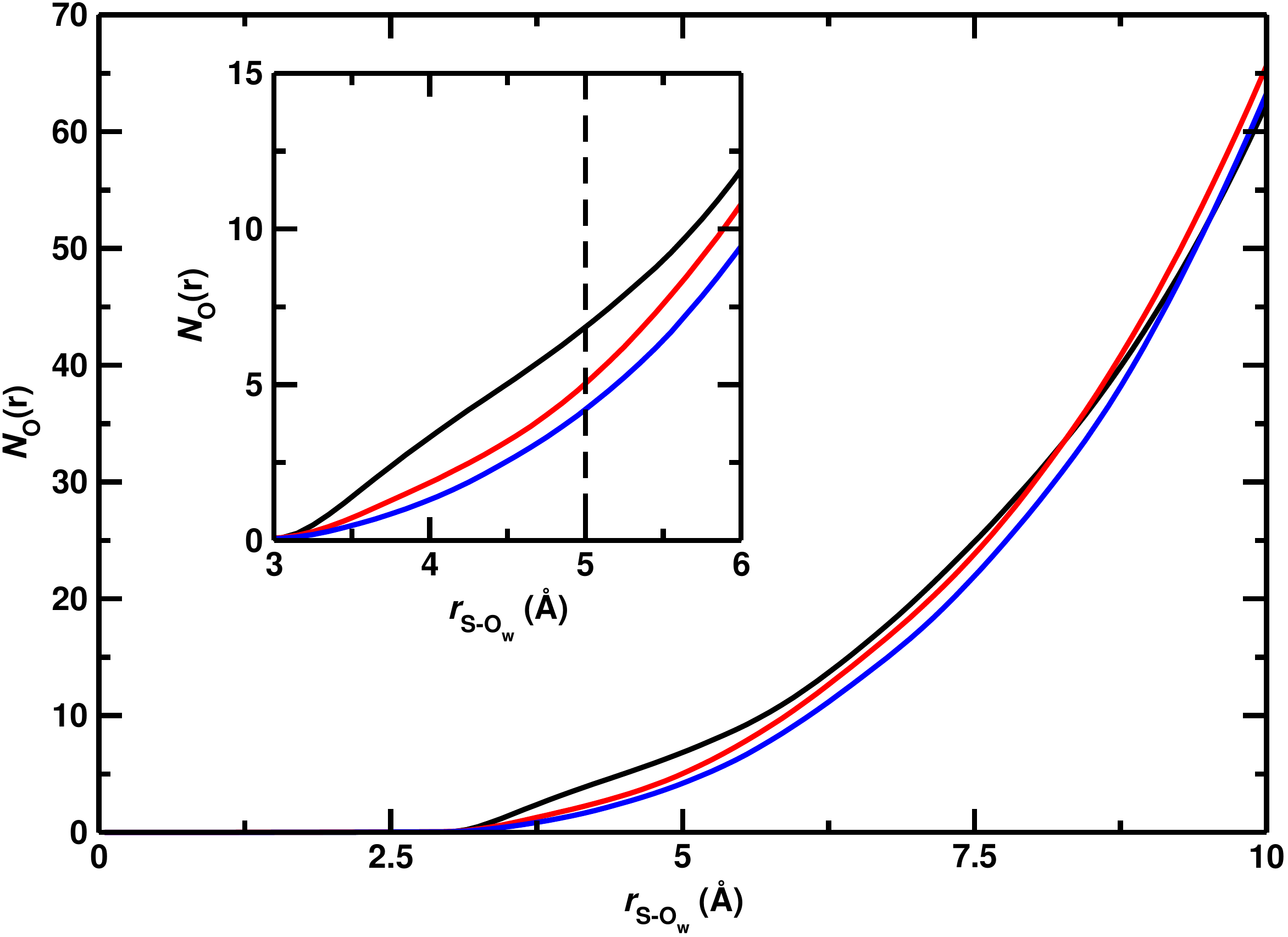}
    \includegraphics[width=0.49\linewidth]{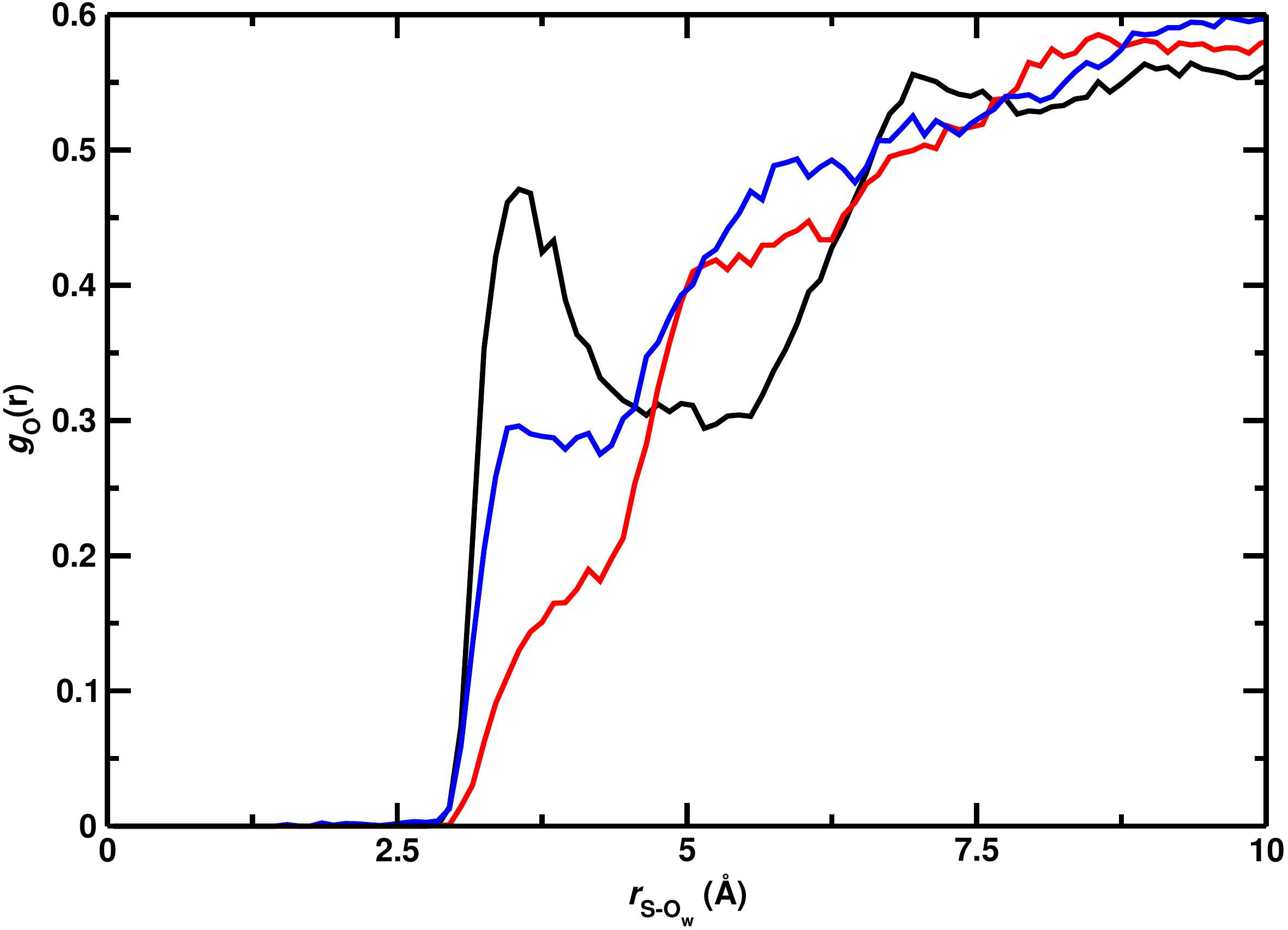}
    \includegraphics[width=0.49\linewidth]{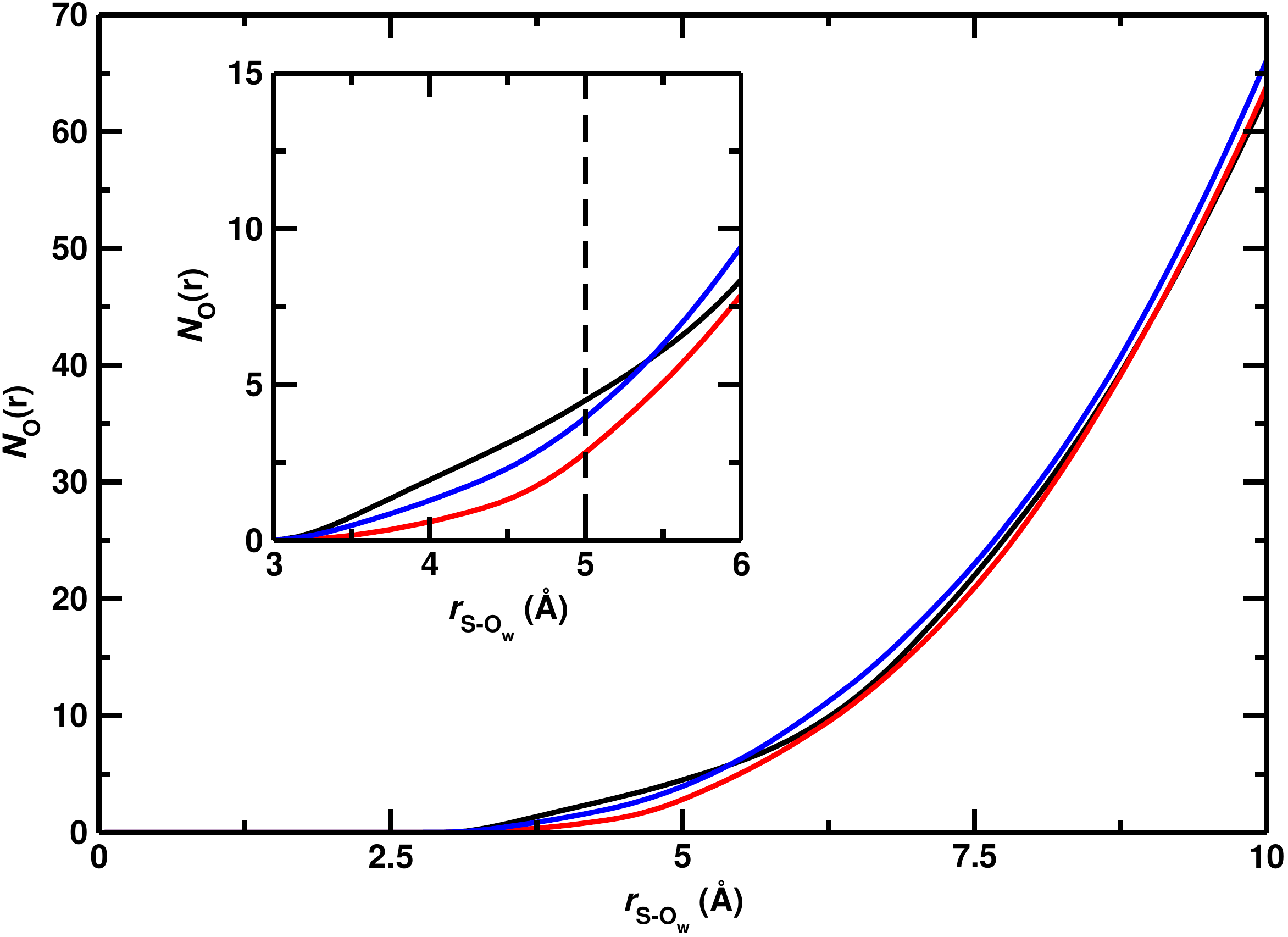}
    \caption{Top Panels: Radial distribution function of water oxygen
      (Left) and the corresponding coordination number of $N_{\rm
        O}(r)$ of water oxygen (Right) with respect to the sulfur atom
      obtained from $NpT$ simulations for KRAS with GDP bound at 300 K
      using point charges. Bottom Panels: For the simulations without
      GDP, the radial distribution function of water oxygen (Left) and
      the corresponding coordination number of $N_{\rm O}(r)$ of water
      oxygen (Right) with respect to the sulfur atom obtained from
      $NpT$ simulations at 300 K with PC model. Color code: WT
      (black), cis-KRASNO (red), trans-KRASNO (blue).}
\label{fig5}
\end{figure}

\noindent
First, the local water ordering around the modification site (Cys118)
for WT and S-nitrosylated KRAS is considered. The radial distribution
function $g_{\rm S-OW}$($r$) and the corresponding number $N_{\rm
  S-OW}$($r$) of water oxygen (OW) with respect to the sulfur atom of
Cys118 in WT, cis-KRASNO, and trans-KRASNO are shown for both,
simulations with and without GDP in the active site, see
Figure~\ref{fig5}. Nitrosylated KRAS remained in its starting cis- and
trans-conformations, respectively, throughout the 10 ns
trajectory. For simulations with GDP at 300 K, hydration around Cys118
substantially differs between WT and cis-, trans-KRASNO in the active
site. Both nitrosylated variants were less hydrated around Cys118 in
the range of $r_{\rm S-OW}$ between 3 and 5 \AA\/ compared with WT. The
first solvation shell peak appears at 3.5 \AA\/ in WT (see
Figure~\ref{fig5}, black line) and cis-KRASNO (red line) whereas no
sharp first solvation peak was observed for trans-KRASNO (blue line).
These results show that the -SNO moiety in the trans-conformation is
less solvent-exposed compared to the cis-conformer. The occupation of
the first solvation shell (up to $r_{\rm S-OW}$ $\sim$ 5) differs by 2
and 3 water molecules, which amounts to 43 and 28\%, with respect to
WT for cis- and trans-KRASNO, respectively.\\

\noindent
For simulations without GDP at 300 K, again, hydration around Cys118
substantially differs between WT and cis-, trans-KRASNO. The
nitrosylated variants were less hydrated in the environment of the
sulfur atom in the range of $r_{\rm S-OW}$ 3--5 \AA\/ compared with
WT. However, contrary to the simulations including GDP, trans-SNO was
more solvent-exposed and thus had increased hydration compared to
cis-SNO. The first solvation shell peak for WT and trans-KRASNO
appears at 3.5 \AA\/. The first solvation shell for cis- and
trans-KRASNO was less occupied by 2 and 1 water molecules compared to
WT which is a reduction of 40 and 20\% for cis- and trans-KRASNO,
respectively. \\

\noindent
The number of water molecules within 5 \AA\/ of the phosphate atoms of
GDP (P$_{\rm GDP}$) in WT, cis- and trans-KRASNO at 300 K as a
function of time is shown in Figure~\ref{sifig3}. The phosphate side
of the GDP selected for the analysis for two reasons: first, electron
rich phosphate atoms engage in electrostatic interactions with water
molecules, hence push or collect water molecules in their vicinity and
secondly, the chemically important part of the GDP where reactions
such as phosphate transfer take place. The hydration around P$_{\rm
  GDP}$ is drastically decreased in cis-KRASNO compared to WT and
trans-KRASNO. The average number of water molecules was 7.2, 3.7 and
7.0 for WT, cis- and trans-KRASNO, respectively. The decrease in
hydration can be rationalized by the increased displacement of the
Switch-I region (Phe28 to Asp38) near GDP in cis-KRASNO compared to
the WT X-Ray structure and WT, trans-KRASNO simulations, (see
Table~\ref{tab1}) which pushed the water molecules away from the
P$_{\rm GDP}$.\\

\begin{figure}[H]
  \centering
  \includegraphics[width=0.49\linewidth]{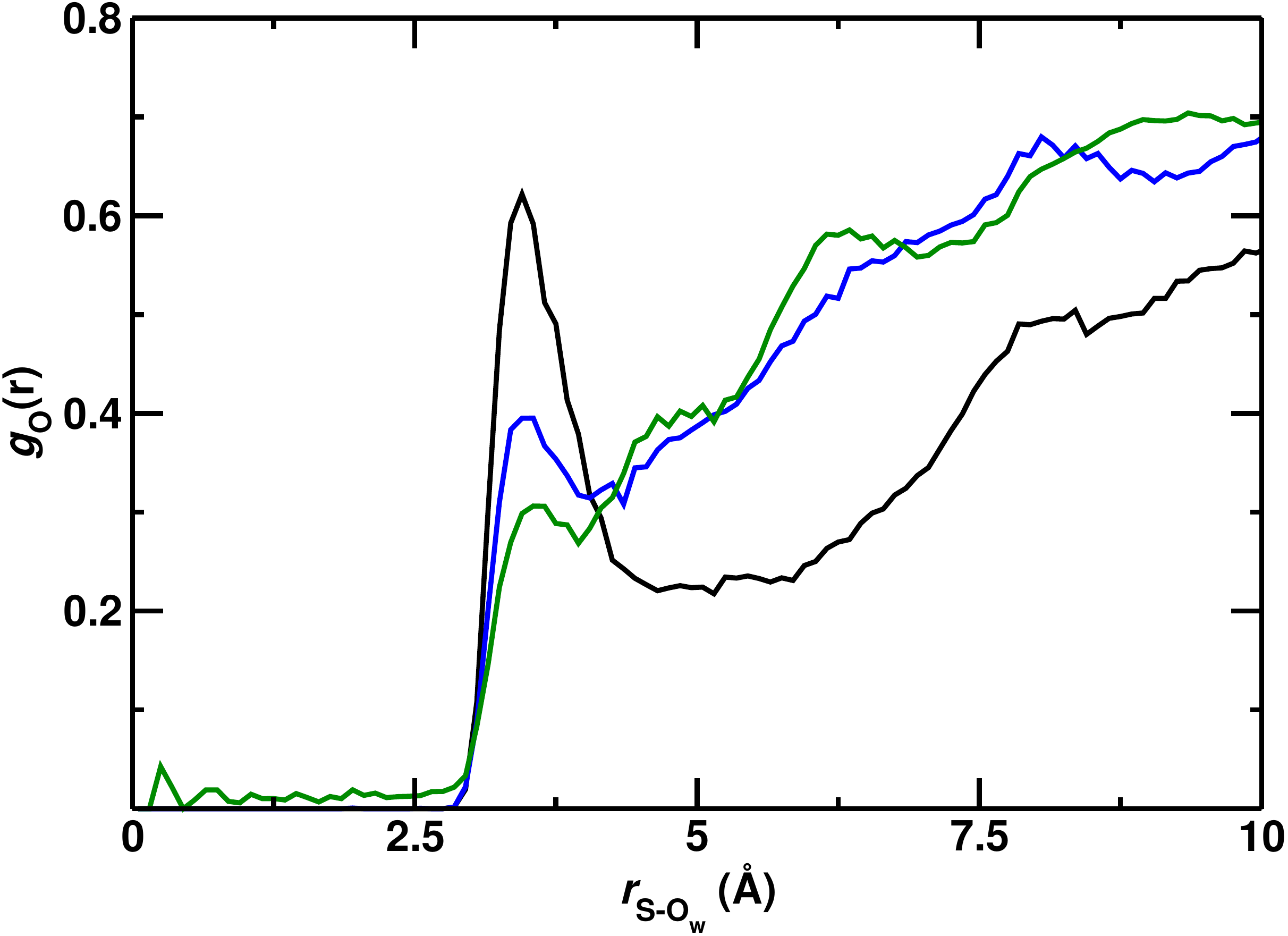}
  \includegraphics[width=0.49\linewidth]{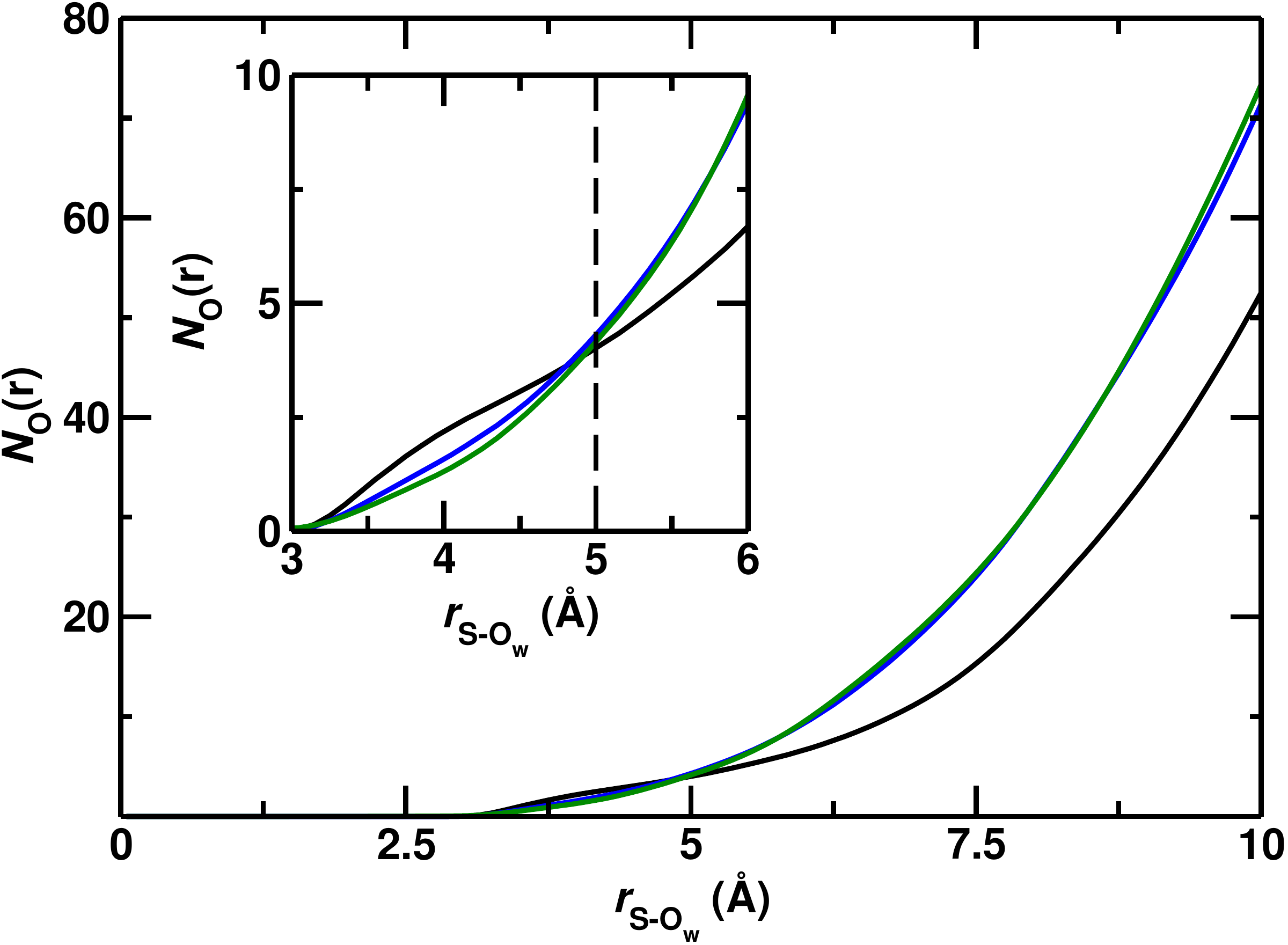}
  \includegraphics[width=0.49\linewidth]{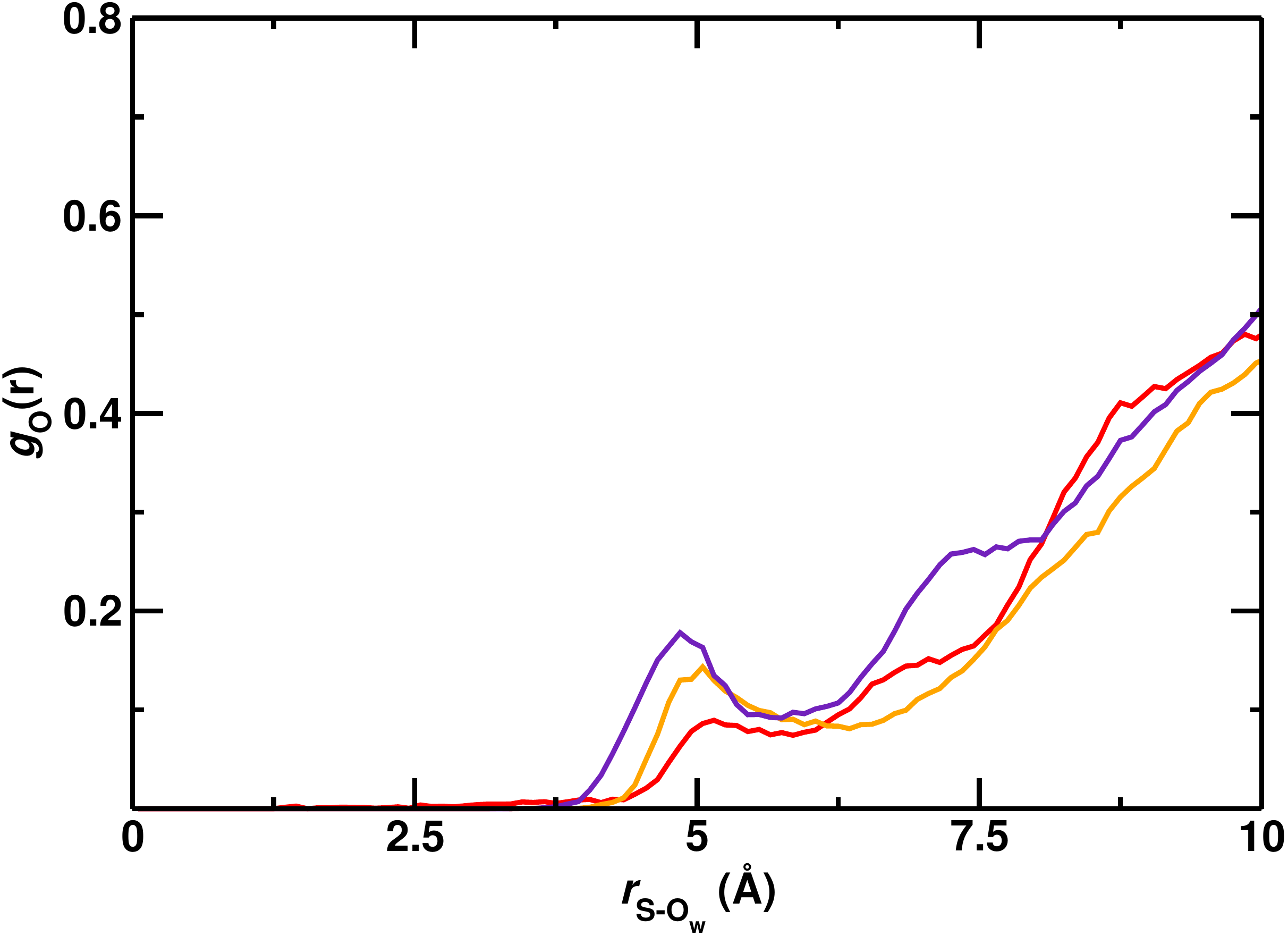}
  \includegraphics[width=0.49\linewidth]{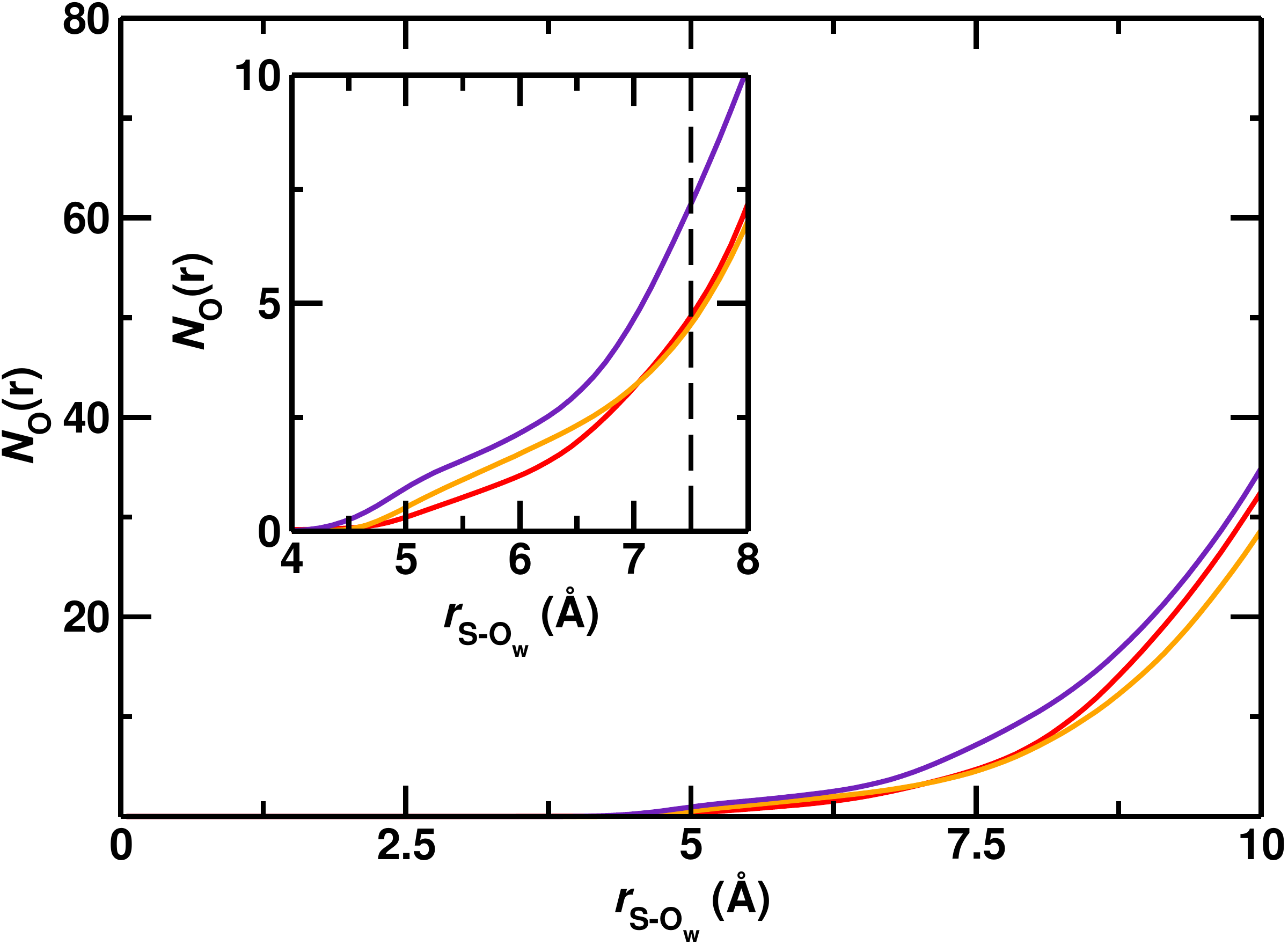}
  \includegraphics[width=0.49\linewidth]{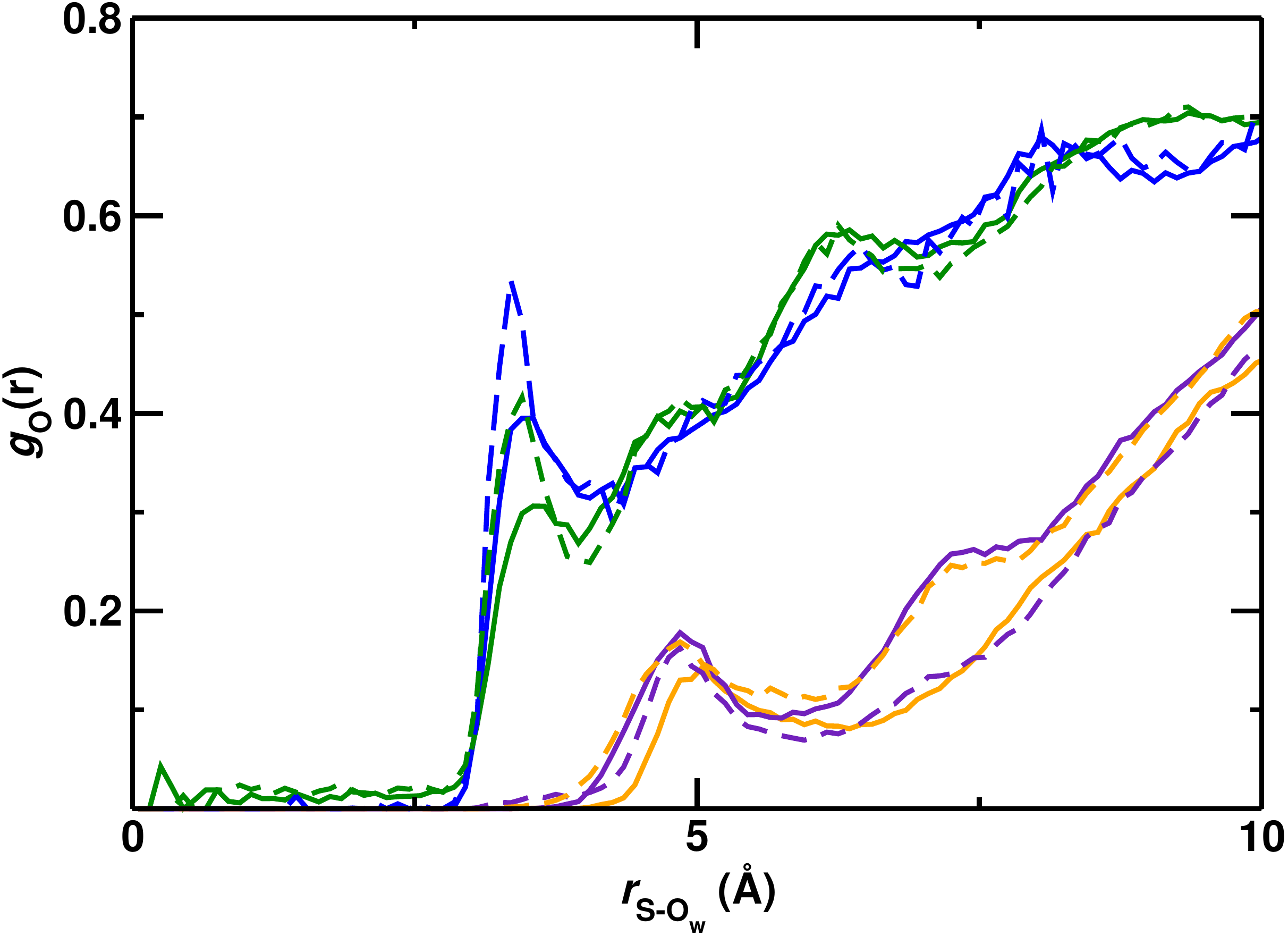}
  \includegraphics[width=0.49\linewidth]{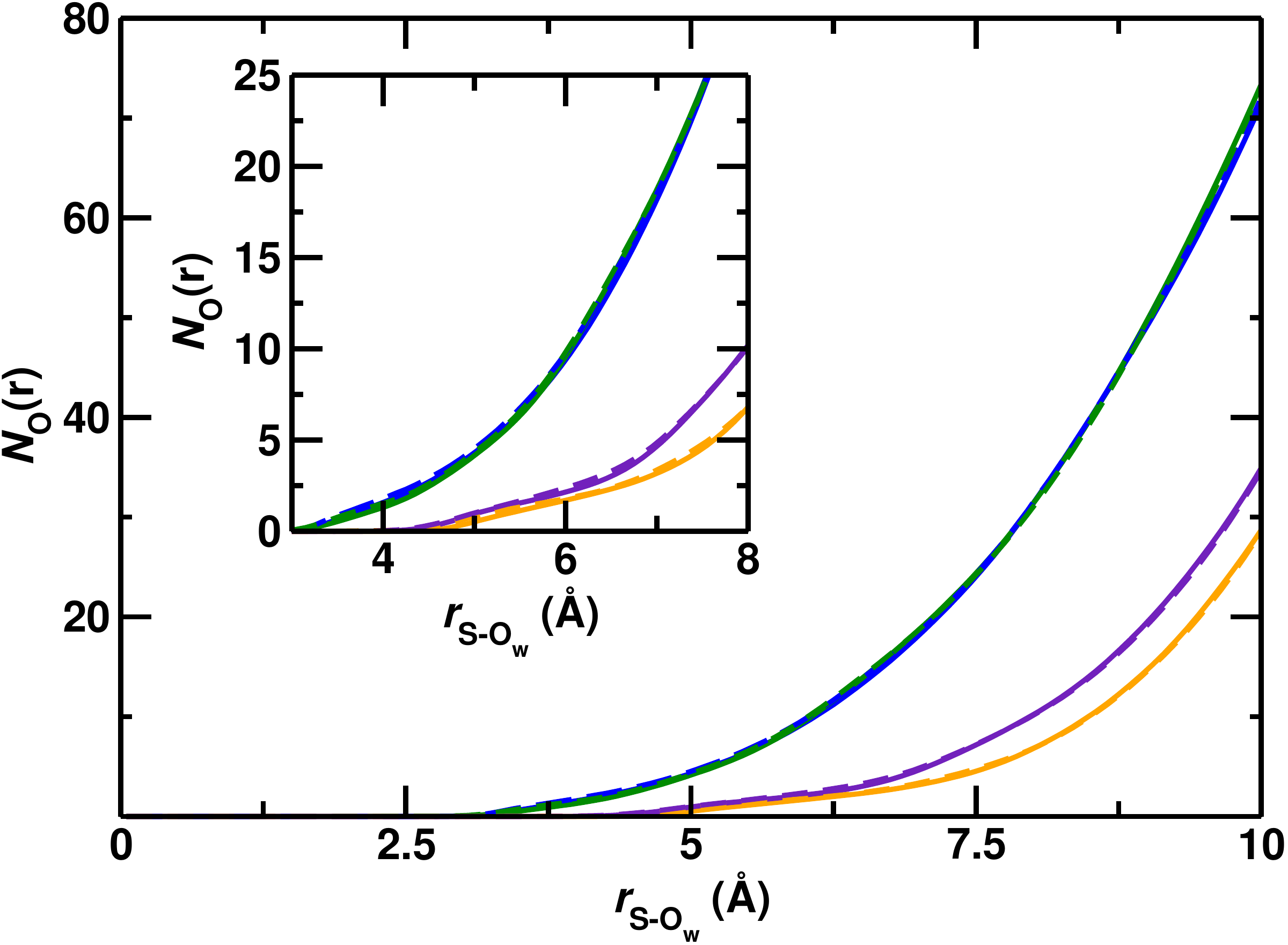}
    \caption{Water structure around Hb, left column for radial
      distribution functions (with respect to sulfur atom of Cys93)
      and right column for total number of water molecules from $NpT$
      simulations at 300 K. Top Panel: T$_0$ (black), cis-T$_0$NO
      (blue), trans-T$_0$NO (green) from PC simulations. Middle
      Panels) R$_4$ (red), cis-R$_4$NO (orange), and trans-R$_4$NO
      (indigo) from PC simulations. Bottom Panels: Solid lines are for
      PC, dashed lines are for MTP. Color code: cis-T$_0$NO (blue),
      trans-T$_0$NO (green), cis-R$_4$NO (orange), and trans-R$_4$NO
      (indigo). }
\label{fig6}
\end{figure}

\noindent
{\it Hemoglobin:} Next, the hydration near the NO-- modification site
(Cys93$\beta$) is considered for WT and S-nitrosylated HbNO is
investigated. The radial distribution function $g_{\rm S-OW}$($r$) and
the corresponding number $N_{\rm S-OW}$($r$) of water oxygen (OW) with
respect to the sulfur atom of Cys93$\beta$ in WT, cis-HbNO, and trans-HbNO
are shown for both T$_0$NO and R$_4$NO states with PC in
Figures~\ref{fig6}. For the T$_0$NO state, hydration at 300 K
decreases significantly after the modification for the first solvation
shell, especially in trans-T$_0$NO. The first solvation shell was
observed at 3.5 \AA\/ for all variants. However, beyond the first
solvation shell, both cis- and trans-T$_0$NO draw more water molecules
around themselves. The same number of water molecules was observed at
5 \AA\/ for all variants, however, for distances larger than 5 \AA\/
number of water molecules increased for the nitrosylated proteins.
The results show that S-nitrosylation pushed waters away in the first
solvation shell but recruit more water after 5 \AA\/ compared to WT.
For the R$_4$NO state, hydration around Cys93$\beta$ was substantially
lowered compared to T$_0$NO. The Cys93$\beta$ was less solvent-exposed in
R$_4$NO and peaks were lower in magnitude. Further, the first
solvation shell peak was observed at 5 \AA\/.  Contrary to T$_0$NO,
the S-nitrosylation deepens the first solvation shell in R$_4$NO for
both cis-, and trans-R$_4$NO. The difference in the total number of
water between WT and modified proteins at 7.5 \AA\/was lower in the
R$_4$NO state.  \\

\noindent
Multipoles up to quadrupoles were also added to the conventional point
charges to determine the effects of more accurate electrostatic model
to the quantification of the local hydration. The radial distribution
function $g_{\rm S-OW}$($r$) and the corresponding number $N_{\rm
  S-OW}$($r$) of water oxygen (OW) with respect to the sulfur atom of
Cys93$\beta$ in cis-HbNO, and trans-HbNO are shown for both T$_0$NO
and R$_4$NO states with PC and MTP in Figures~\ref{fig6}. For the
T$_0$NO state, the local hydration increased with MTP model for both
cis- and trans-T$_0$NO compared to PC. The first solvation shells
around 3 \AA\/ were deeper. However, the difference between PC and MTP
is diminished after 4.5 \AA\/. For R$_4$NO state, the effect of
additional multiples was marginal compared to PC. The number of water
molecules around the Cys93$\beta$ did not change between the models.\\

\noindent
Small peaks at 0 to 2.75 \AA\/ before the first solvation shell,
stating at 2.75 \AA\/ were observed for the $g_{\rm O}(r)$ of
trans-T$_0$NO, which is not present in other proteins, see
Figures~\ref{fig6}. To further characterize these peaks, a total
number of waters 2.75 \AA\/ around the Cys93$\beta$ as a function of
time is computed, and presented in Figure~\ref{sifig4}. The total
number of water molecules ranged from 1 to 12 throughout the
trajectory. The substantial changes in the total number indicate a
migration (or diffusion) of water. Especially, a decrease in hydration
after 8 ns was present. On average 5 water molecules were around 2.75
\AA\/ of Cys93$\beta$.\\

\subsection{Local Hydrophobicity Analysis}
In this section, the local hydrophobicity (LH) around the
S-nitrosylated Cys118 for KRAS and Cys93$\beta$ for Hb proteins are
investigated. LH ($\delta \lambda_{phob}^{r}(t)$) is time-dependent
quantity to determine and quantify solvent exposure of a given amino
acid residue $r$ at time $t$. LH is a quantity that can be considered
as a prolongation of the radial distribution function, which
encompasses both orientation of a given water molecule and its
distance from the interface into consideration. In this study, the
probability density function of $\delta \lambda_{phob}^{r}(t)$ is
present and positive LH values ($\delta \lambda_{phob}^{r} >$ 0)
considered as indicative for hydrophilicity.\\

\begin{figure}[H]
  \centering
  \includegraphics[width=0.49\linewidth]{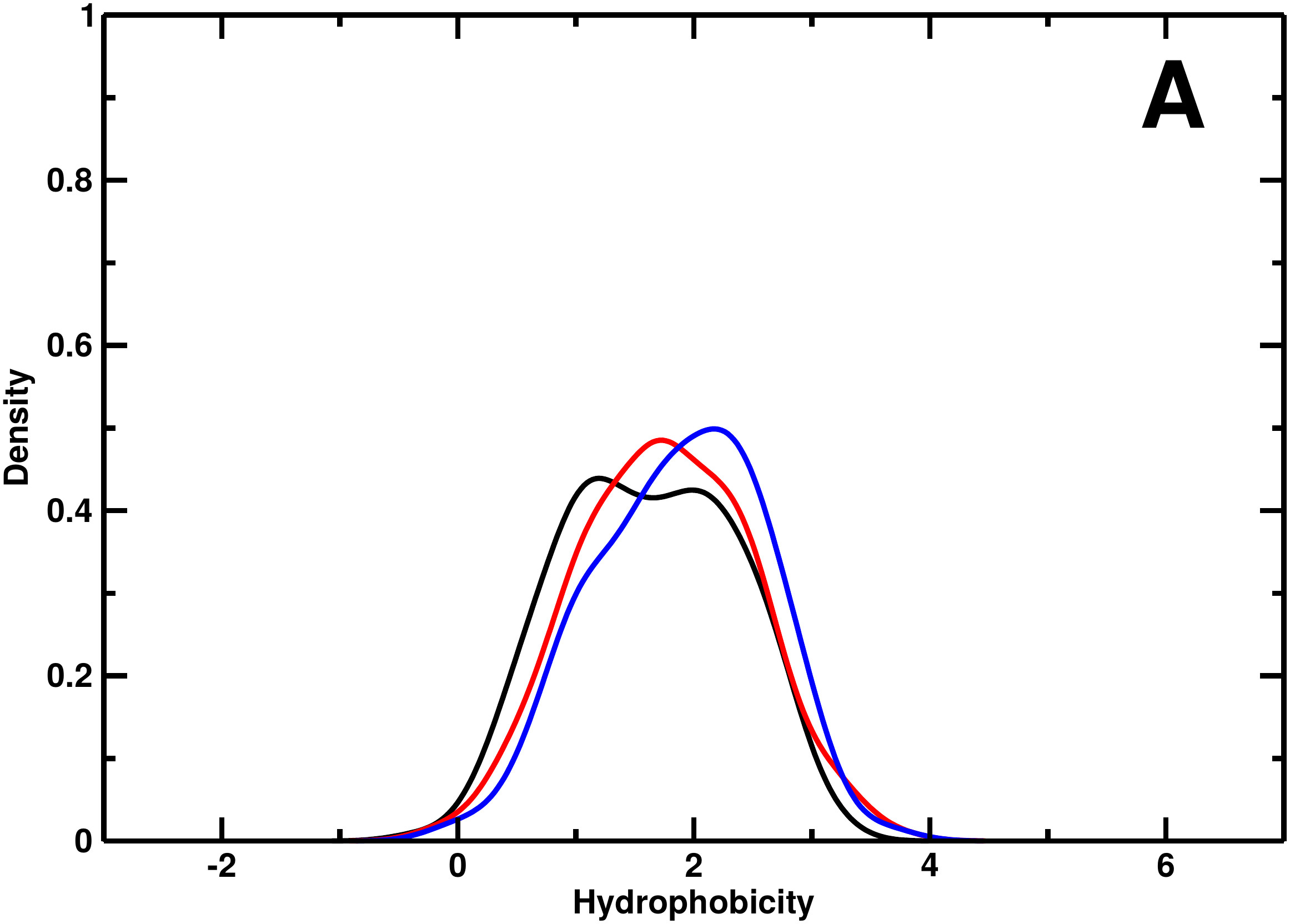}
  \includegraphics[width=0.49\linewidth]{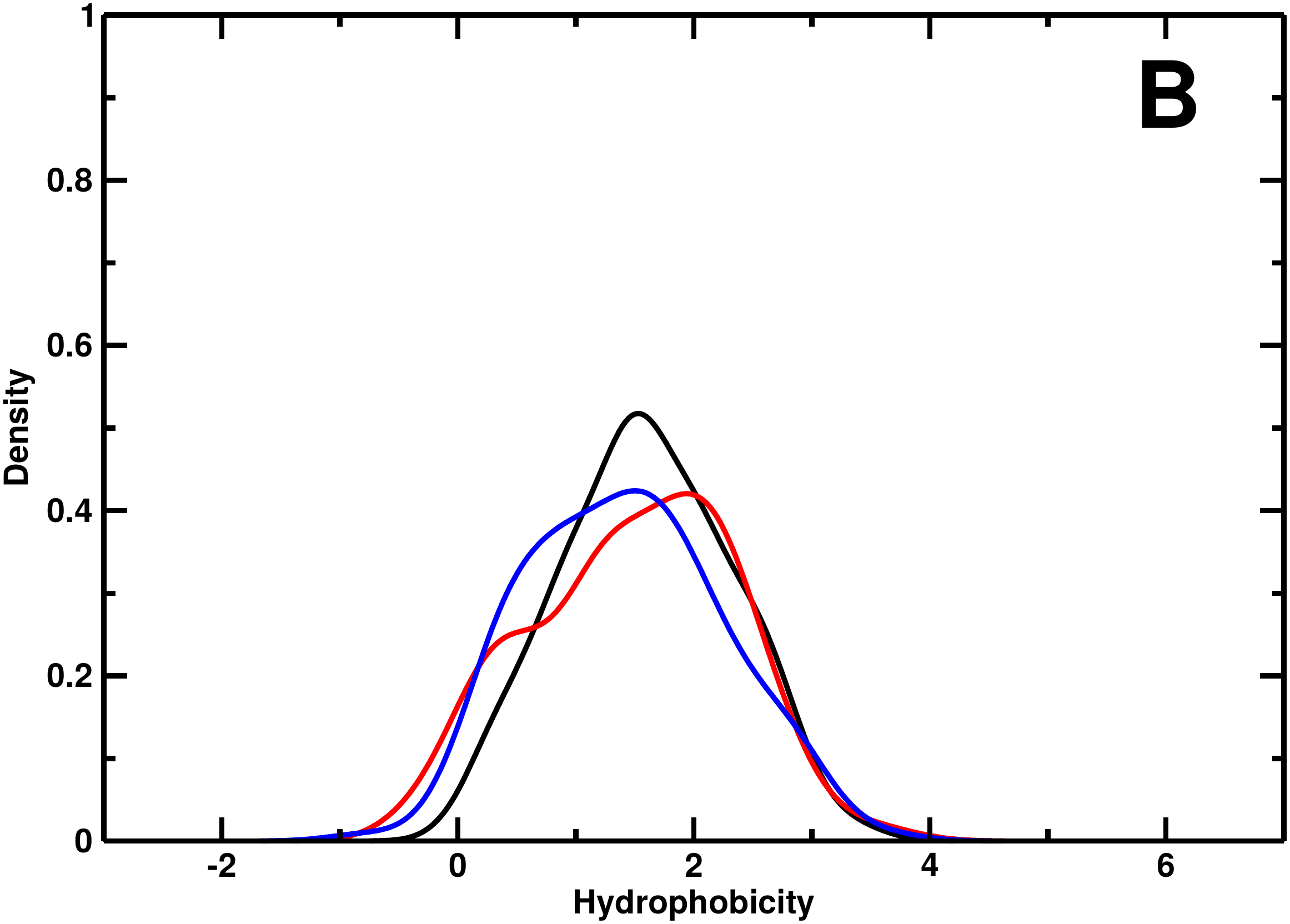}
  \includegraphics[width=0.48\linewidth]{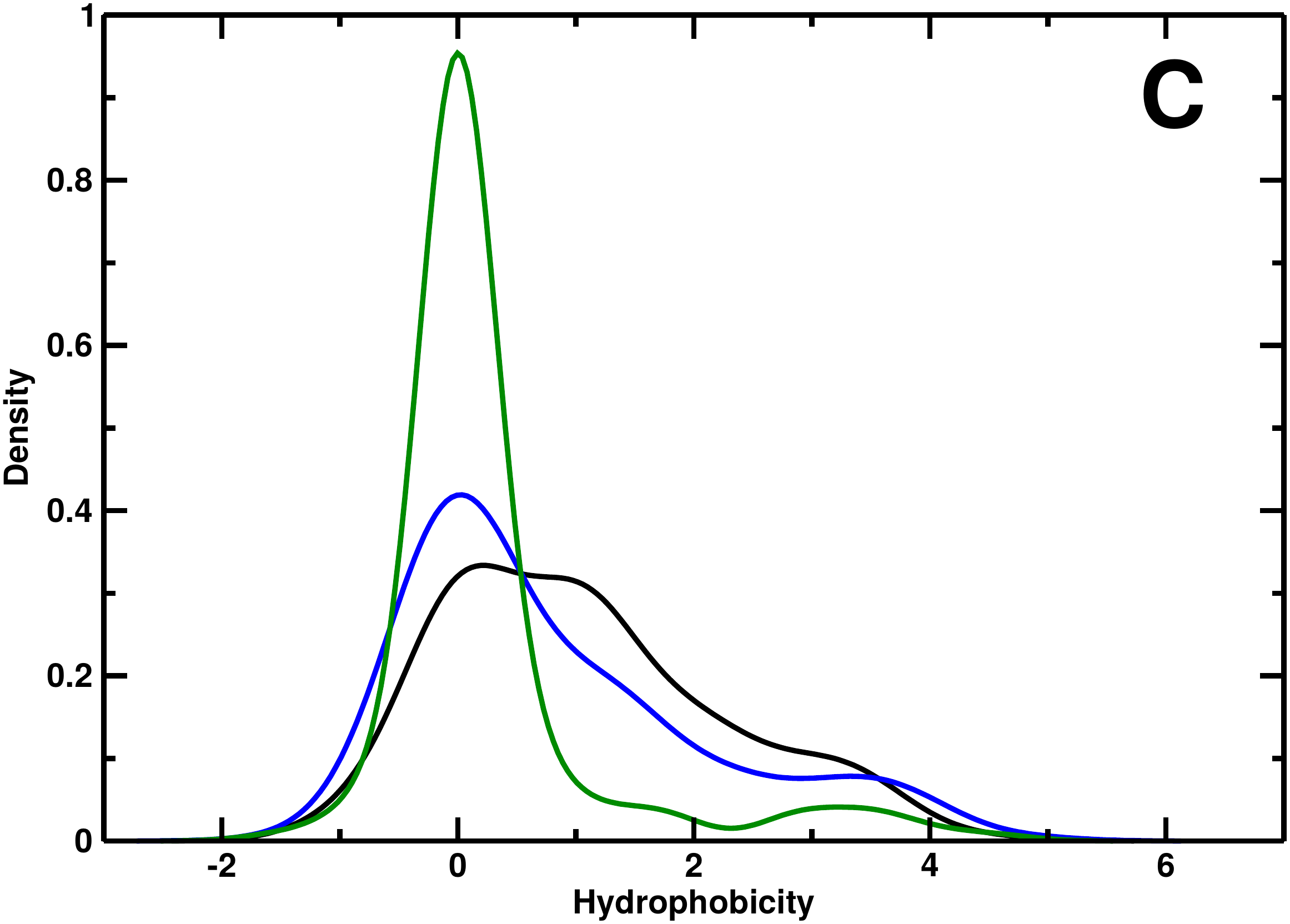}
  \includegraphics[width=0.49\linewidth]{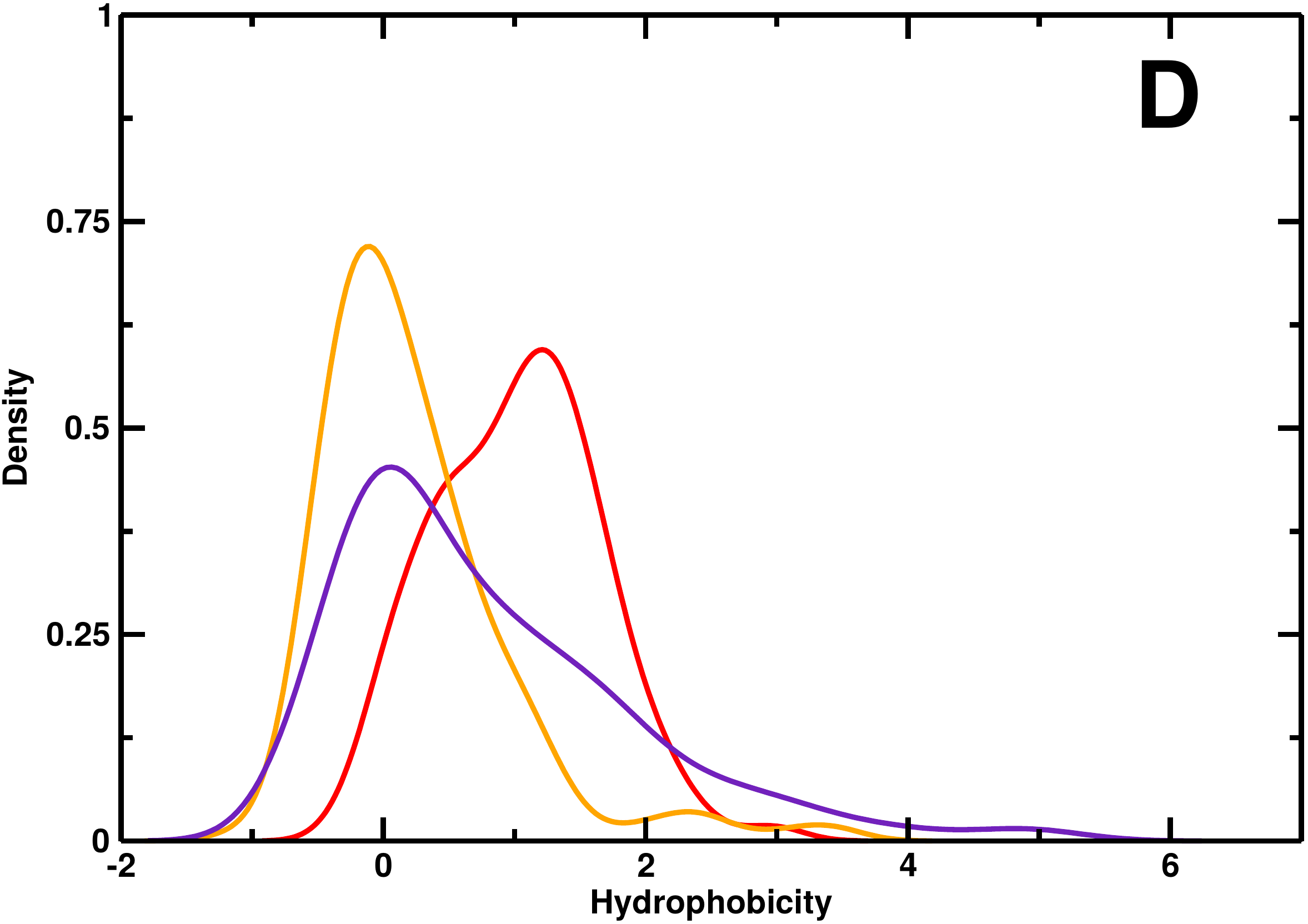}
    \caption{Panel A: The distribution of hydrophobicity values of
      Cys118 from 10 ns $NpT$ simulations at 300 K for WT KRAS
      (black), cis-KRASNO (red), and trans-KRASNO (blue). Panel B: The
      distribution of hydrophobicity values of Cys118 from 10 ns $NpT$
      simulations at 300 K for WT KRAS without GDP (black), cis-KRASNO
      without GDP (red), and trans-KRASNO without GDP (blue). C) The
      distribution of hydrophobicity values of Cys93$\beta$ from 10 ns
      $NpT$ simulations at 300 K for T$_{\rm 0}$ (black), cis-T$_{\rm
        0}$NO (blue), and trans-T$_{\rm 0}$NO (green). D) The
      distribution of hydrophobicity values of Cys93$\beta$ from 10 ns
      $NpT$ simulations at 300 K for R$_{\rm 4}$ (red), cis-R$_{\rm
        4}$NO (orange), and trans-R$_{\rm 4}$NO (indigo) }
\label{fig7}
\end{figure}

\noindent
{\it KRAS:} The distribution of hydrophobicity values of Cys118 for WT
KRAS (black), cis-KRASNO (blue), and trans-KRASNO (green) is shown in
Figure~\ref{fig7}A, and WT KRAS without GDP (black), cis-KRASNO
without GDP (blue), and trans-KRASNO without GDP (green) shown in
Figure~\ref{fig7}B. For the simulations with GDP (see Panel A), the
hydrophobicity distributions were similar for all KRAS variants. They
had a broad density peak for LH between [0,4] which emphasizes the
strong hydrophilic character of Cys118 residue. The highest density
points for cis-KRASNO were at 1.8 whereas 2.1 for trans-KRASNO. The WT
KRAS had multimodal distribution with density maxima at 1 and 2.1. For
simulations without GDP (see Panel B), again all variants showed
strong hydrophilic character with a broad density peak between
[0,4]. The WT KRAS without GDP had monomodal distribution, whereas
cis-KRASNO had a shoulder around 0. The highest density points were
between [1.9, 2.1] for all variants. \\

\noindent
{\it Hemoglobin:} The distribution of hydrophobicity values of
Cys93$\beta$ for T$_{\rm 0}$ (black), cis-T$_{\rm 0}$NO (blue),
trans-T$_{\rm 0}$NO (green), R$_{\rm 4}$ (red), cis-R$_{\rm 4}$NO
(orange), and trans-R$_{\rm 4}$NO (indigo) shown in
Figure~\ref{fig7}. The T$_{\rm 0}$ (Panel A) had the highest density
in [0, 1.5] regions and had positive skew distribution which means
T$_{\rm 0}$ had mostly hydrophilic character. Although S-nitrosylated
variants also showed some hydrophilic characters, the highest density
values for both cis- and trans- conformers were closer to 0.  The
cis-T$_{\rm 0}$NO had positive skew distribution similar to T$_{\rm
  0}$, the highest density observed at 0. However, the positive skew
emphasizes the hydrophilic character. The trans-T$_{\rm 0}$NO had the
highest density around in [-1, 1] with a second density maximum around
3, which means it shows mostly hydrophobic character throughout the
simulation compared to T$_{\rm 0}$.  \\

\noindent
For R$_{\rm 4}$ (Panel B), the density maximum was at 1.7, and the
distribution had a shoulder around 0.5. Although it had negative skew,
distributed values were mostly positive emphasizing a strong
hydrophilic character. The density maxima for the S-nitrosylated
variants were at 0, and both had positive skews. Both cis- and trans-
conformers showed less hydrophilic character compared to R$_{\rm
  4}$. A similar trend was also observed in T$_{\rm 0}$ proteins.  \\

\subsection{Infrared Spectroscopy in the -NO Stretch Region}
\begin{figure}[H]
  \centering \includegraphics[width=0.49\linewidth]{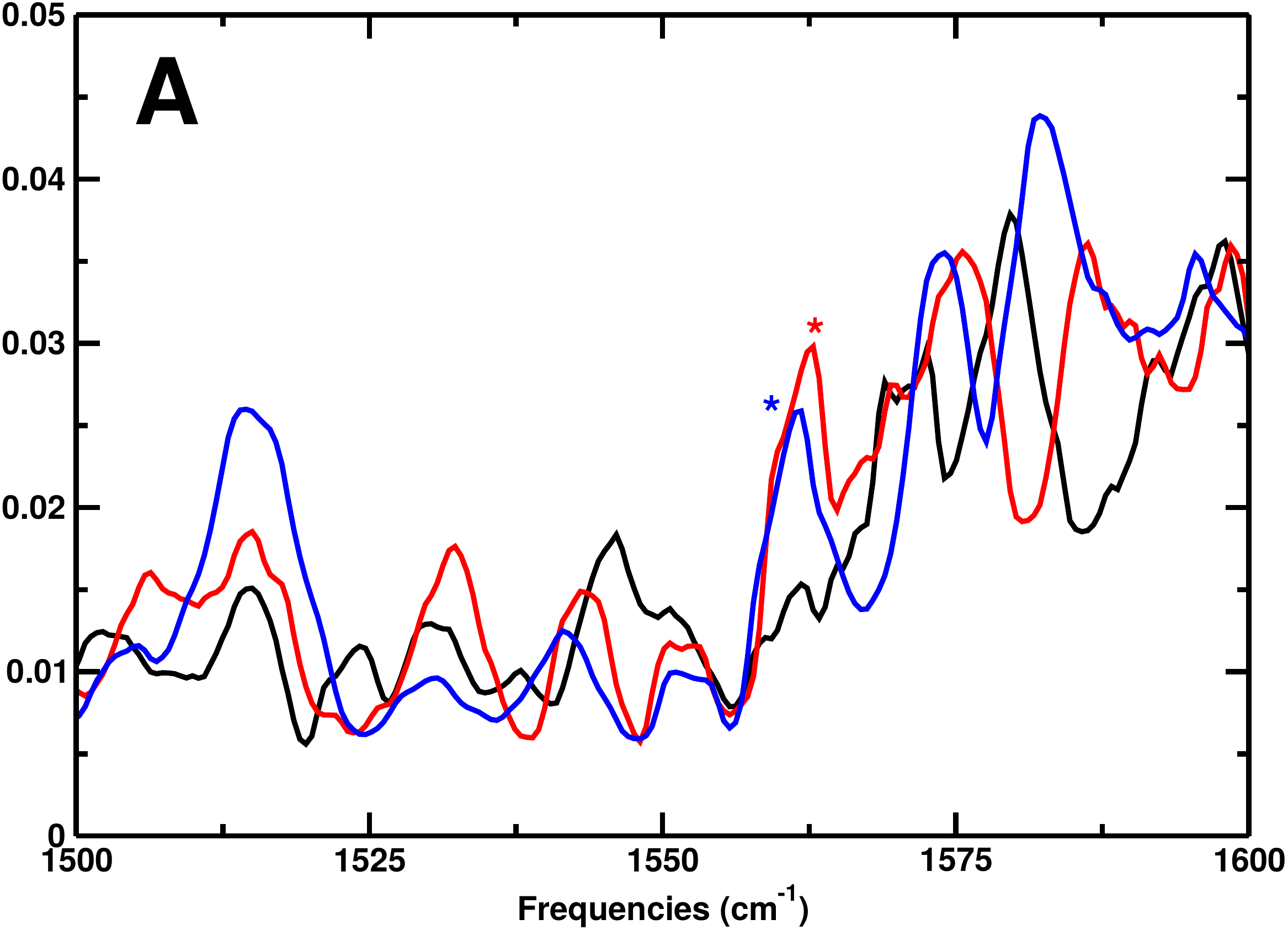}
  \includegraphics[width=0.49\linewidth]{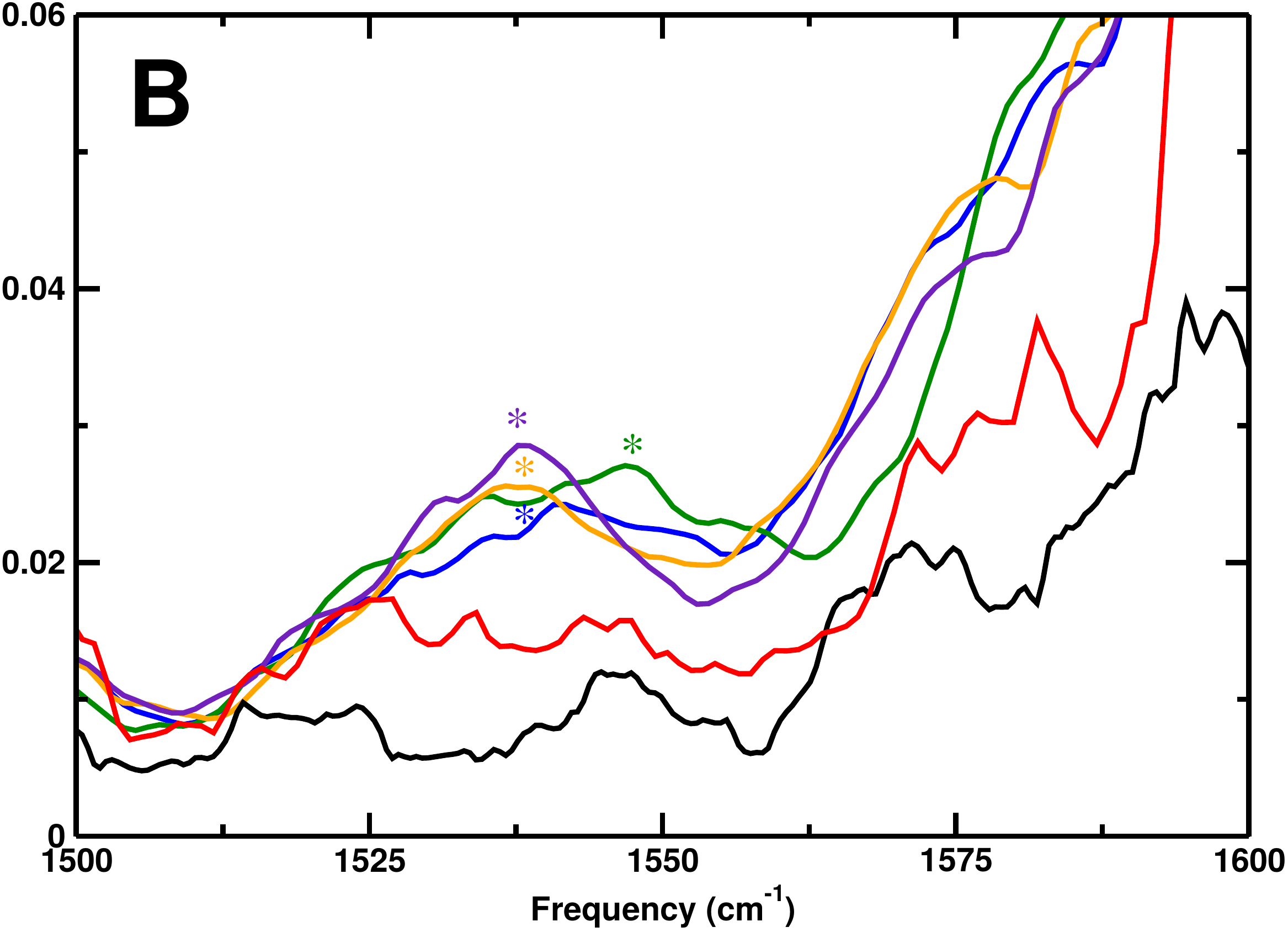}
    \caption{Panel A: IR spectra from the total protein dipole moment
      time series for WT without GDP (black), cis-KRASNO with GDP
      (red), and trans-KRASNO with GDP at 50 K. The $^{14}$N$^{18}$O
      isotope is used for the cis- and trans-KRASNO. Panel B: IR
      spectra for Hb, T$_0$ (black), R$_4$ (red), cis-T$_0$NO (blue),
      trans-T$_0$NO (green), cis-R$_4$NO (orange), and trans-R$_4$NO
      (indigo). Spectra are range between 1500 and 1600 cm$^{-1}$
      presented. The colored stars label the spectral signatures to
      which the mode was assigned based on analysis of the power
      spectra.}
\label{fig8} 
\end{figure}

One direct way to detect whether or not nitrosylation occurred, is by
the use of infrared spectroscopy. The IR and power spectra for
nitrosylated KRAS and Hb are presented in Figures~\ref{fig8} and
\ref{sifig5}, respectively. \\
           
\noindent
{\it KRAS:} For KRAS, determining the IR band due to the NO stretch
for the natural isotope was not possible due to congested spectra in
the amide I region. Thus, the IR spectra related to the
$^{14}$N$^{18}$O stretch are reported in Figure~\ref{fig8} for WT
(black), cis- and trans-KRASNO (red, blue) with GDP at 50 K. Such a
low temperature was required to avoid washing out the peak that
belongs to the NO-stretch. For the $^{14}$N$^{18}$O stretch, the peak
for cis-KRASNO from the simulations appears at 1561 cm$^{-1}$ compared
with 1563 cm$^{-1}$ for trans-KRASNO. Also, the peak intensity was
higher for the trans variant with respect to cis. The assignment of
these peaks is also possible by considering the power spectra
corresponding to the NO stretch motion. The power spectra of
cis-KRASNO with $^{14}$N$^{16}$O (black), $^{14}$N$^{18}$O (red) and
$^{15}$N$^{18}$O (blue). trans-KRASNO with $^{14}$N$^{16}$O (green),
$^{14}$N$^{18}$O (yellow) and $^{15}$N$^{18}$O (magenta) has shown in
Figure~\ref{sifig5}. The identical simulations for all three isotopic
variants lead to red-shifts of the $^{14}$N$^{18}$O-stretch by
[--27,--20] cm$^{-1}$ for [cis,trans]-KRASNO and
$^{15}$N$^{18}$O-stretch by [--59,--61] cm$^{-1}$ for
[cis,trans]-KRASNO compared to the natural isotope
$^{14}$N$^{16}$O. The red-shifts observed for the isotopic
substitution are on par with previous
examples.\cite{turan2021spectroscopy} \\

\noindent
{\it Hemoglobin:} For Hb, the NO peak was observed with the natural
isotope, and IR spectra related to NO stretch are reported in
Figure~\ref{fig8} for cis-T$_0$NO, trans-T$_0$NO, cis-R$_4$NO, and
trans-R$_4$NO at 50 K. The peak for [cis,trans]-T$_0$NO appears at
[1540,1547] cm$^{-1}$ whereas for [cis,trans]-R$_4$NO appears at
[1537,1538]. The results show that R$_4$NO variants are red-shifted by
3 and 9 cm$^{-1}$ compared to cis- and trans-T$_0$NO,
respectively. Further, trans- conformers were blue-shifted by 7 and 1
cm$^{-1}$ compared to cis- conformers for both T$_0$NO and
R$_4$NO. The shifts show that cis, trans- conformers, especially for
T$_0$NO, and T$_0$NO, R$_4$NO proteins can be distinguished with IR
spectra.  \\

\section{Conclusions}
The present work reports on the structural, dynamical, hydration, and
spectroscopic implications of nitrosylation of cysteine residues. For
this, S-nitrosylated KRAS without and with GDP, and the T$_0$ and
R$_4$ structural substates of Hb were considered. The overall finding
is that nitrosylation influences local hydration, and the local and
global dynamics of the proteins.  More specifically, for KRAS it was
found that attaching NO to Cys118 rigidifies the Switch-I region which
has functional implications because this region acts as a binding
interface for effector proteins and RAS regulators. On the other hand,
for Hb in its T$_0$ and R$_4$ substates the flexibility of secondary
structural motives was increased following S-nitrosylation. These
effects were confirmed from analyzing DCCM maps which imply that
correlated movements between the switch-I region and nearby helix C
increase for KRASNO. Similarly, for Hb the amplitude of correlated
motions increased for T$_0$NO compared with R$_4$NO.\\

\noindent
For local hydration it is found that S-nitrosylation in KRAS decreases
water access by up to 40 \%. Also, cis- versus trans-orientations of
the modifications lead to different local hydration. This is also
found for T$_0$NO and R$_4$NO with the additional observation that for
the R$_4$ conformation the hydration is yet lower than for the T$_0$NO
substate. Finally, the infrared spectra in the NO stretch region were
determined and were found to exhibit characteristic peaks and shifts
depending the ligation state (for KRAS) or conformational
substate. However, the absorption occurs in a spectrally congested
region of the protein-IR spectrum which will make experimental
detection of these bands difficult.\\

\noindent
In conclusion, for KRAS and Hb considered in the present work the
simulations find characteristic differences between WT and
nitrosylated variants of the proteins with regards to local hydration
and local and global flexibility of the proteins.l This, together with
earlier finding for Mb, suggests that attaching NO to cysteine
residues at the surface of proteins has functional implications.\\

\section*{Supporting Information}
The supporting information provides RMSD and power spectra.

\section{Acknowledgment}
This work was supported by the Swiss National Science Foundation
grants 200021-117810, 200020-188724, the NCCR MUST, and the University
of Basel which is gratefully acknowledged.

\bibliography{refs}

\clearpage

\renewcommand{\thetable}{S\arabic{table}}
\renewcommand{\thefigure}{S\arabic{figure}}
\renewcommand{\thesection}{S\arabic{section}}
\renewcommand{\d}{\text{d}}
\setcounter{figure}{0}  
\setcounter{section}{0}  

\noindent
{\bf SUPPORTING INFORMATION: Local Hydration Control and Functional
  Implications Through S-Nitrosylation of Proteins: Kirsten rat
  sarcoma virus (KRAS) and Hemoglobin (Hb)}

\begin{figure}[H]
  \centering
    \includegraphics[width=0.6\linewidth]{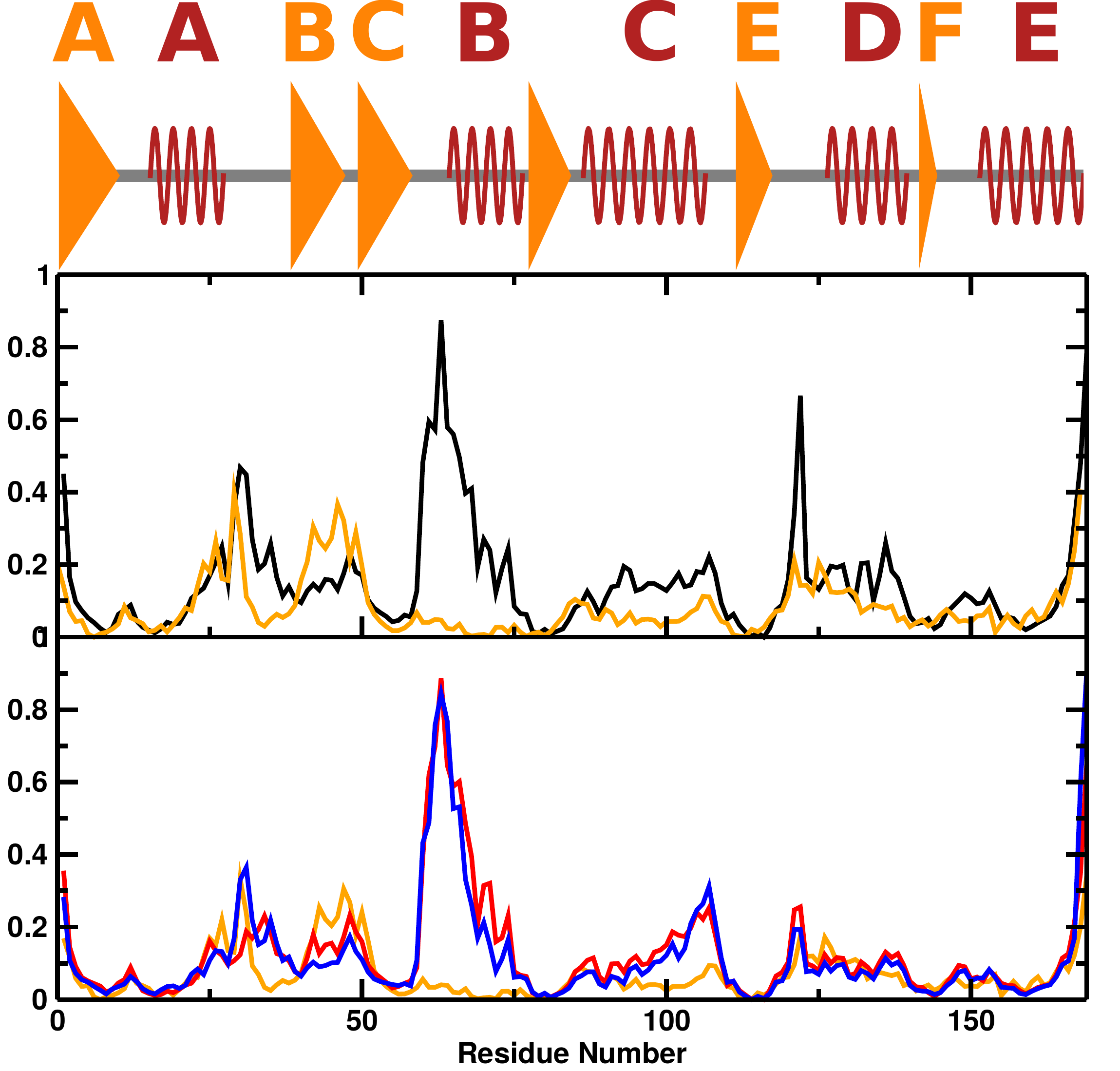}
    \caption{Superposition of the experimentally measured C$_{\alpha}$
      B-factors and the computed C$_{\alpha}$ RMSFs from the present
      simulations. Up) Experimental WT B-Factors (orange) vs. WT
      (black) RMSF at 300 K. Bottom) WT B-Factors vs cis- (red) and
      trans-KRASNO (blue) RMSF at 300 K. B-Factor and RMSF values are
      scaled and normalized. Orange triangles indicate the position of
      $\beta$-sheets and red helices indicate the position of
      $\alpha$-helices}
\label{sifig1}
\end{figure}

\begin{figure}[H]
  \centering
    \includegraphics[width=0.49\linewidth]{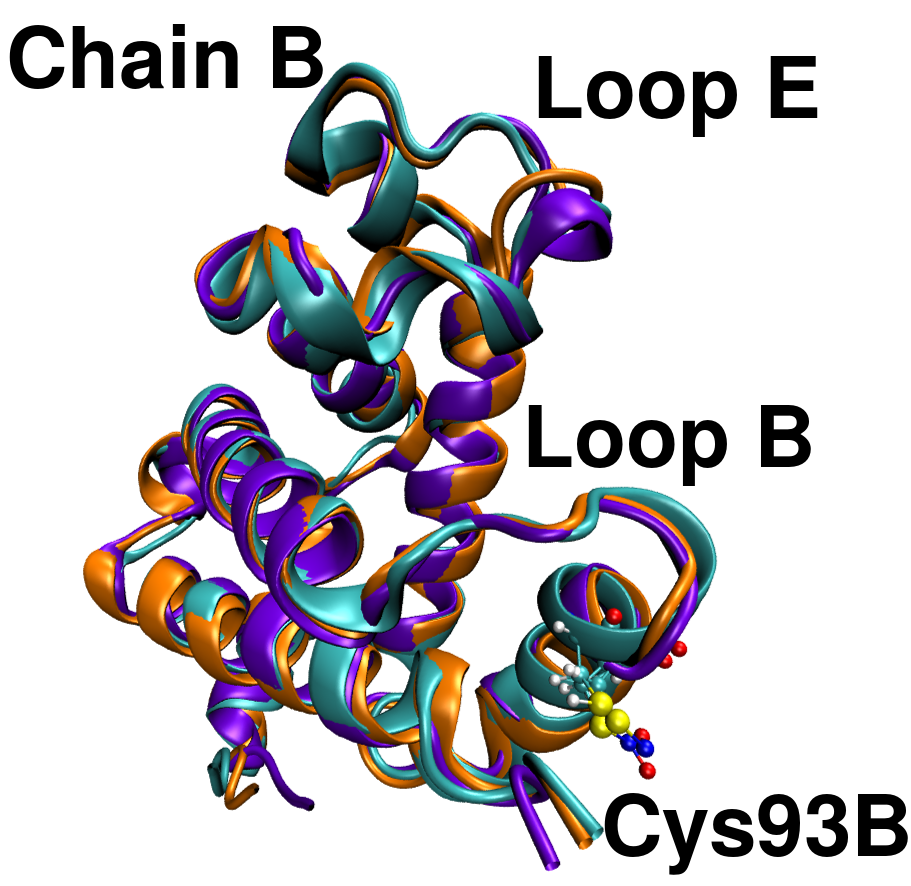}
    \includegraphics[width=0.49\linewidth]{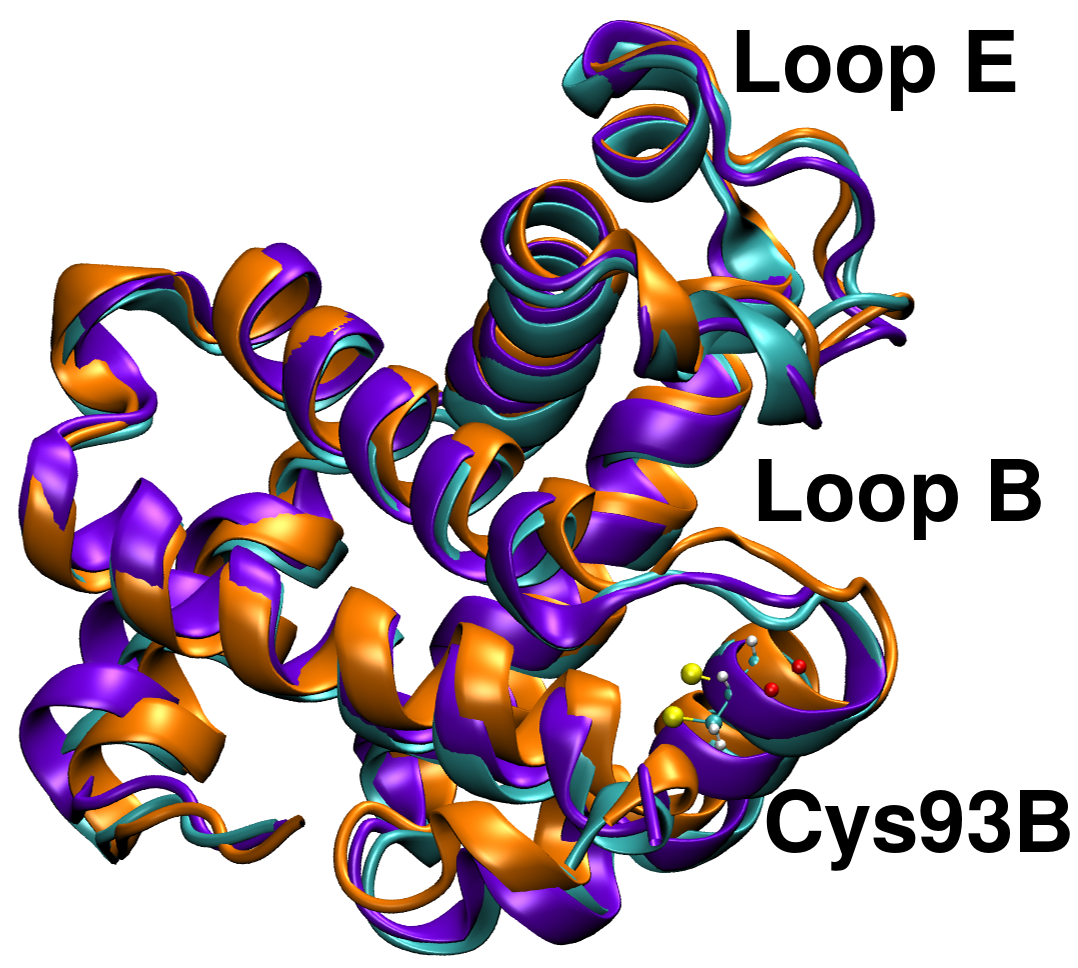}
    \caption{Conformational changes for Hb induced by S-nitrosylation
      at Cys93$\beta_1$ from simulations at 300 K. Left structurel:
      T$_{0}$ (cyan), cis-T$_{0}$NO (orange), trans-T$_{0}$NO
      (indigo). Right structure: R$_{4}$ (cyan), cis-R$_{4}$NO
      (orange), trans-R$_{4}$NO (indigo) S-nitrosylayed Cys93$\beta$
      is represented by CPK.}
\label{sifig2}
\end{figure}

\begin{figure}[H]
  \centering \includegraphics[width=0.49\linewidth]{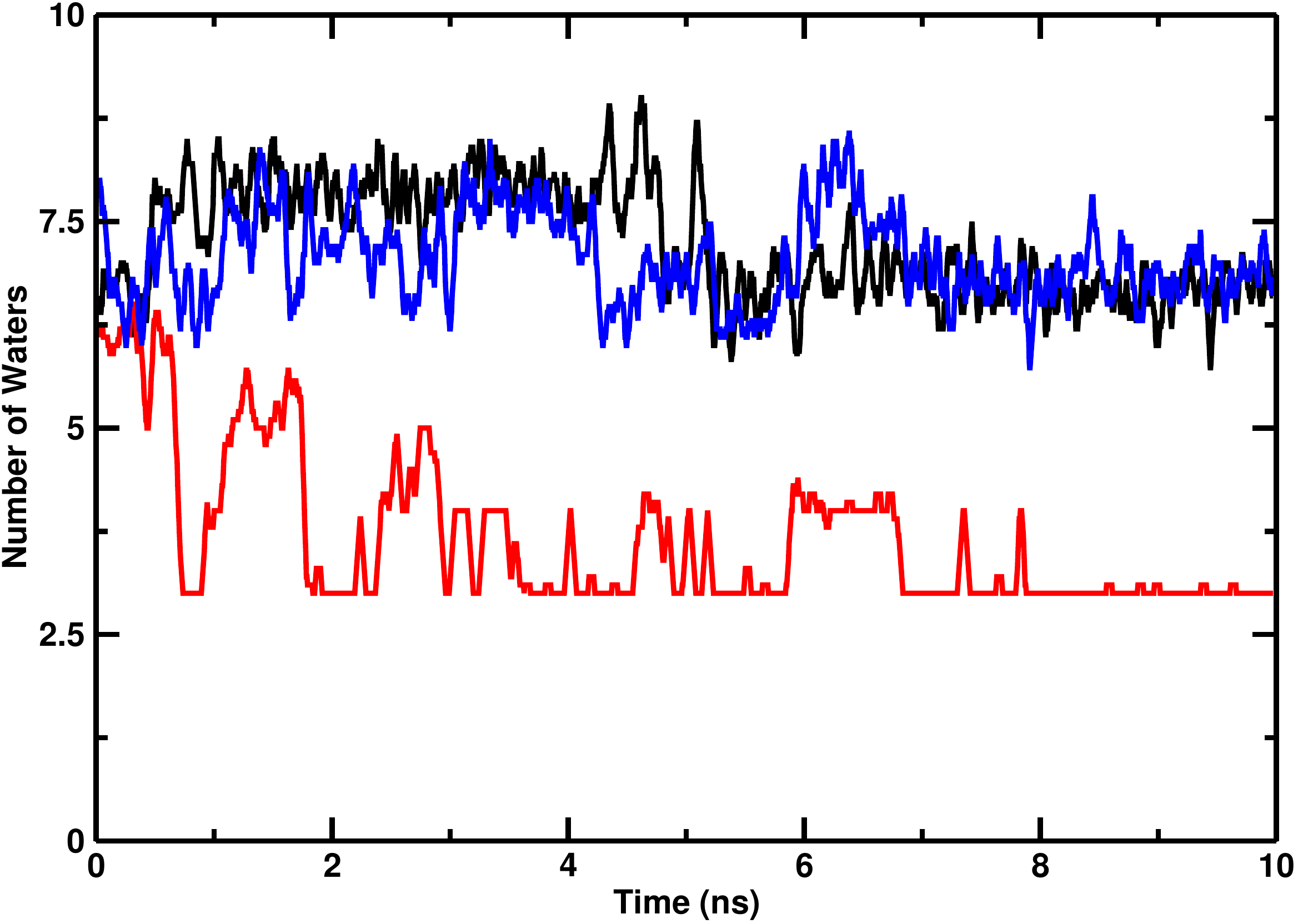}
    \caption{The number of water molecules within 5 \AA\/ of the
      phosphate atoms of GDP in WT (black), cis- (red) and
      trans-KRASNO (blue) at 300 K as a function of time.}
\label{sifig3}
\end{figure}

\begin{figure}[H]
  \centering
    \includegraphics[width=0.49\linewidth]{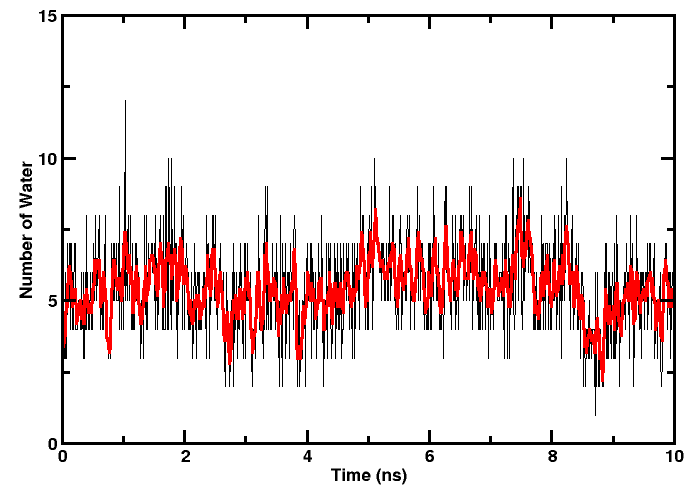}
    \caption{The number of solvent water molecules within 2.75 \AA\/
      of the Cys93B in trans-T$_0$NO at 300 K from simulations with
      point charges.}
\label{sifig4}
\end{figure}

\begin{figure}[H]
  \centering
    \includegraphics[width=0.6\linewidth]{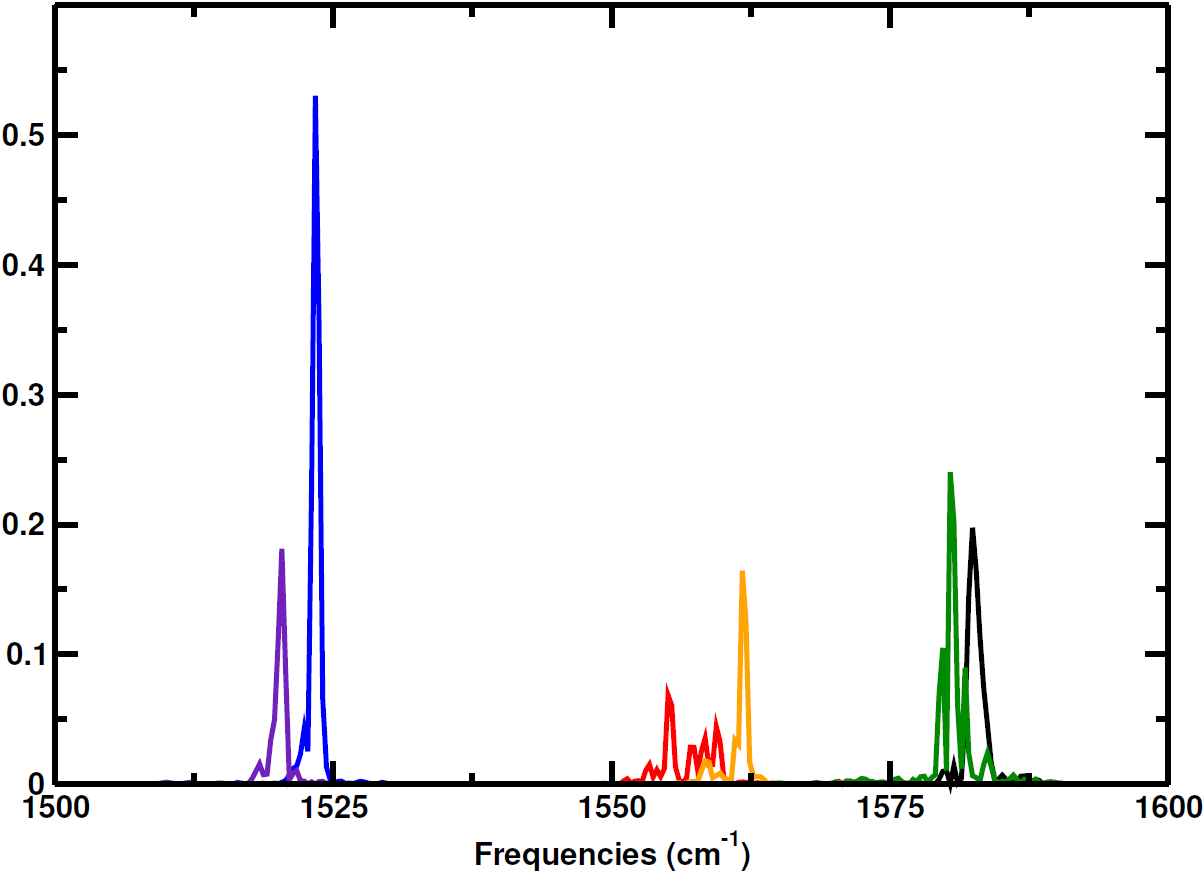}
    \caption{The power spectra for cis- and trans-KRASNO at 50
      K. Color code: cis-KRASNO with $^{14}$N$^{16}$O (black),
      $^{14}$N$^{18}$O (red) and $^{15}$N$^{18}$O (blue). trans-KRASNO
      with $^{14}$N$^{16}$O (green), $^{14}$N$^{18}$O (orange) and
      $^{15}$N$^{18}$O (indigo).}
\label{sifig5}
\end{figure}

\end{document}